\documentclass{elsart}
\usepackage{rotating}
\usepackage{graphicx}
\usepackage{xspace}

\usepackage{amssymb}

\begin{document}

\begin{flushright}
Fermilab-Pub-06-007-AD\\
 BNL-76776-2006-JA\\
\end{flushright}

\begin{frontmatter}




\title{Secondary Beam Monitors\\ for the NuMI Facility at FNAL}

\author[ut]{S. Kopp\corauthref{myemail}}
\corauth[myemail]{Corresponding author e-mail {\tt kopp@mail.hep.utexas.edu} } 
\author[bnl]{M. Bishai}
\author[bnl]{M. Dierckxsens}
\author[bnl]{M. Diwan}
\author[uw]{A.R. Erwin}
\author[fnal]{D.A. Harris}
\author[ut]{D. Indurthy}
\author[ut]{R. Keisler}
\author[ut]{M. Kostin}
\author[ut]{M. Lang}
\author[pitt]{J. MacDonald}
\author[fnal]{A. Marchionni}
\author[ut]{S. Mendoza}
\author[fnal]{J. Morfin}
\author[pitt]{D. Naples}
\author[pitt]{D. Northacker}
\author[ut]{\v{Z}. Pavlovi\'{c}}
\author[ut]{L. Phelps}
\author[uw]{H. Ping}
\author[ut]{M. Proga}
\author[uw]{C. Vellissaris}
\author[bnl]{B. Viren}
\author[ut]{R. Zwaska}

\address[bnl]{Brookhaven National Laboratory, Upton, Long Island, NY }
\address[fnal]{Fermi National Accelerator Laboratory, Batavia, IL 60510}
\address[pitt]{Department of Physics, University of Pittsburgh, Pittsburgh, PA 15260} 
\address[ut]{Department of Physics, University of Texas, Austin, TX 78712}
\address[uw]{Department of Physics, University of Wisconsin, Madison, WI 53706}

\begin{abstract}
The Neutrinos at the Main Injector (NuMI) facility is a conventional neutrino beam which produces muon neutrinos by focusing a beam of mesons into a long evacuated decay volume.  We have built four arrays of ionization chambers to monitor the position and intensity of the hadron and muon beams associated with neutrino production at locations downstream of the decay volume.  This article describes the chambers' construction, calibration, and commissioning in the beam.  
\end{abstract}

\begin{keyword} neutrino beam; neutrino, pion, muon detectors; ionization chambers; particle sources and targets; beam characteristics; 

\PACS 13.15.+g, 95.55.Vj, 29.40.Cs, 29.25.-t, 29.27.Fh
\end{keyword}
\end{frontmatter}

\section{Introduction}
\label{beamon_intro}

\begin{sidewaysfigure}[p]
  \centering
  \includegraphics[width=23cm]{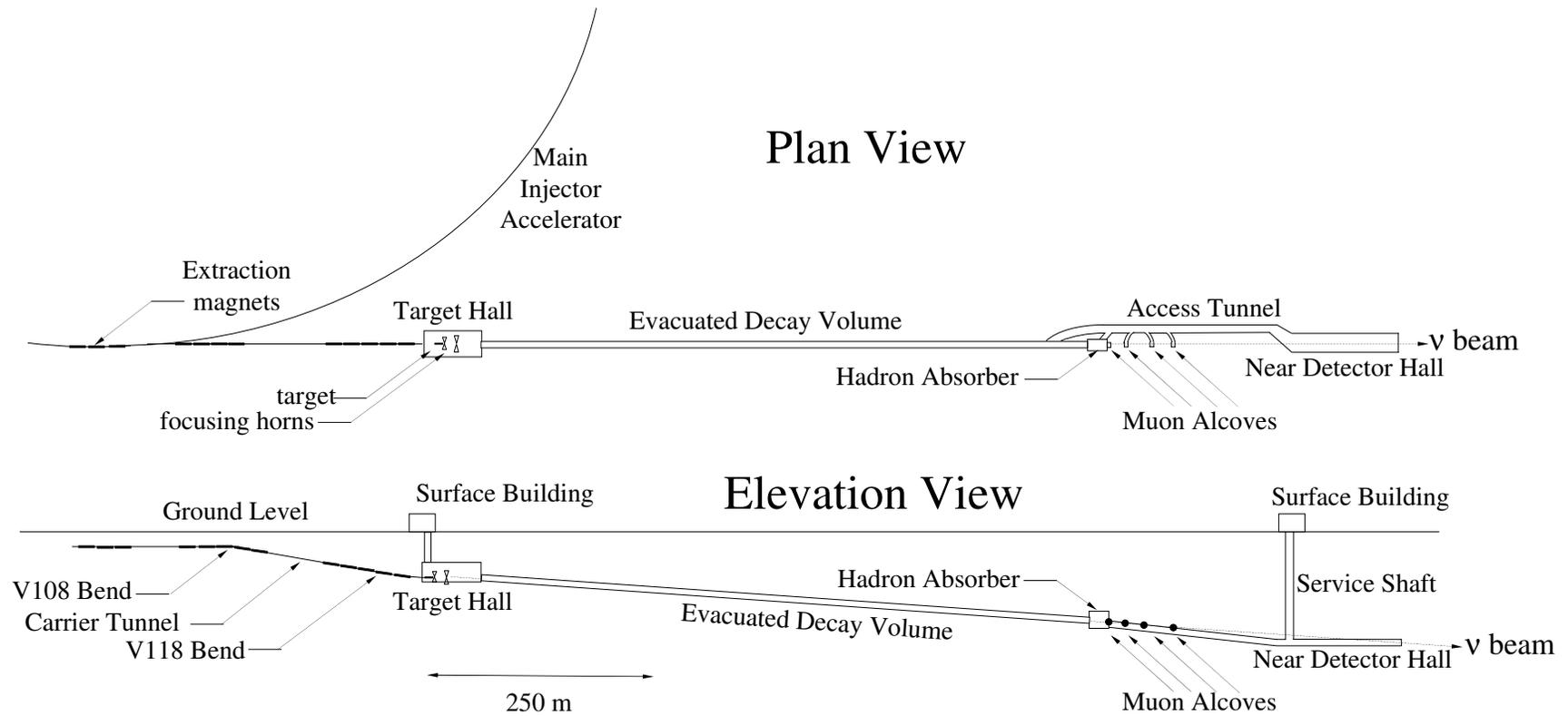}
\vskip .5 cm
  \caption{Plan and elevation views of the NuMI beam facility.  A proton beam is directed onto a target, where the secondary pions and kaons are focused into an evacuated decay volume via magnetic horns.  Ionization chambers at the
end of the beam line measure the secondary hadron beam and tertiary muon beam.}
  \label{fig:numi}
\end{sidewaysfigure}

\begin{figure}[t]
  \centering
\vskip -.5cm
  \includegraphics[width=5. in]{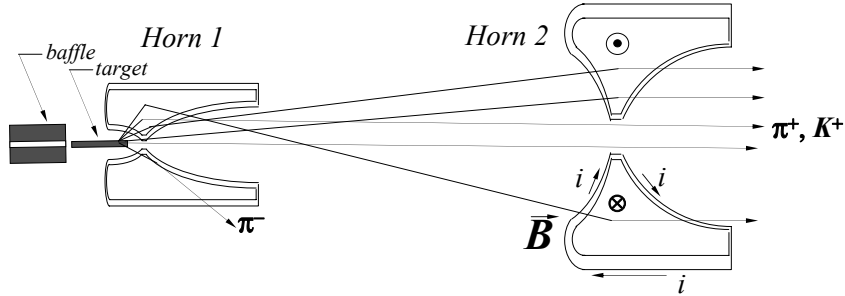}
  \caption{NuMI two-horn beam:  Horns 1 and 2 are separated by 10~m.  A collimating baffle upstream of the target protects the horns from direct exposure to errant proton beam pulses.  The target and baffle system can be actuated further upstream of the Horns to produce higher energy neutrino beams \cite{flexybeam}.  Note that the vertical scale is $4\times$ that of the horizontal (beam axis) scale.}
  \label{fig:targ-horns}
\end{figure}

The Neutrinos at the Main Injector (NuMI) beam line \cite{numitdr,kopp-numi} at the Fermi National Accelerator Laboratory (FNAL) delivers an intense muon neutrino  beam to the MINOS \cite{minostdr} detectors at FNAL and at the Soudan Laboratory in Minnesota.  Additional experiments \cite{nova,minerva} are being planned.  A schematic diagram of the NuMI beam line is shown in Figure~\ref{fig:numi}.
The primary proton beam is fast-extracted from the 120~GeV Main Injector accelerator onto the NuMI pion production target.  The beam line is designed to accept up to $4\times10^{13}$~protons-per-pulse (ppp) with a repetition rate of 0.53~Hz.  After the graphite target, two toroidal magnets called "horns" sign-select and focus the secondary mesons from the target (pions and kaons), as shown in Figure~\ref{fig:targ-horns}.  The mesons are directed into a 675 m long volume, evacuated to $\sim0.5$~Torr to reduce pion absorption, where they may decay to muons and neutrinos.  At the end of the decay volume, a beam absorber stops the remnant hadrons, followed by approximately 240~m of unexcavated rock which attenuates the tertiary muons, leaving only neutrinos.  

\begin{figure}[b]
  \centering
  \includegraphics[width=5. in]{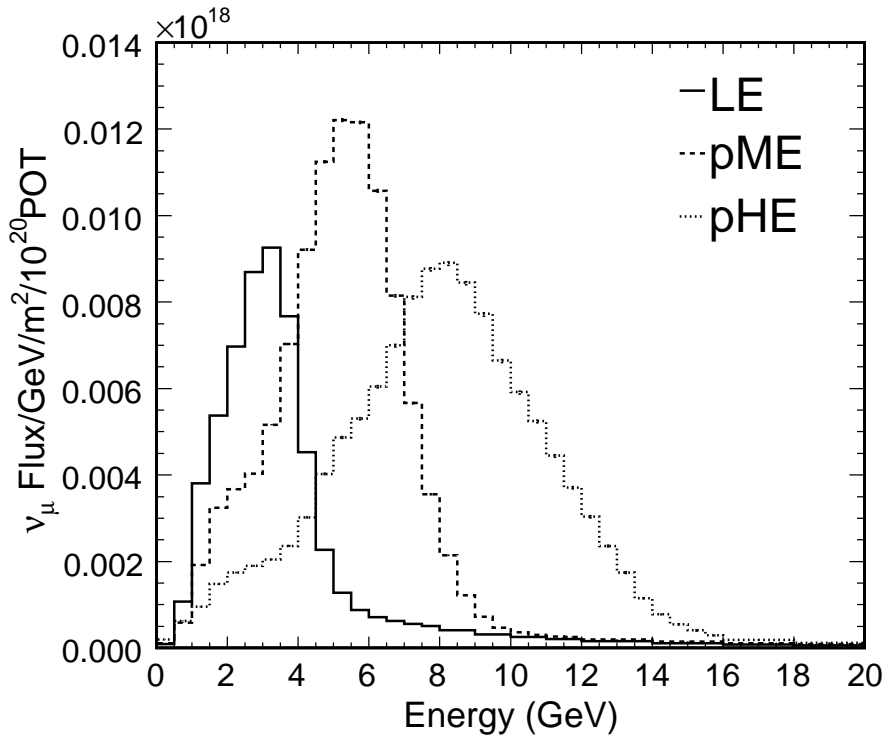}
  \caption{Calculated flux of muon neutrinos in the detector hall located 1040~m from the NuMI target.  Three spectra are shown, corresponding to the low, medium, and high neutrino energy positions of the target \cite{flexybeam} (the ``LE'', ``pME, and ``pHE'' configurations).  In these configurations, the target is located 10, 100, and 250~cm upstream of its fully-inserted position.}
  \label{fig:numi-spectra}
\end{figure}

The target may be positioned remotely so as to produce a variety of wide band beams with peak energies ranging from 3~GeV to 9~GeV \cite{flexybeam}.  The target, shown fully-inserted into the first focusing horn in Figure~\ref{fig:targ-horns}, is mounted on a rail drive system and can be driven as much as 2.5~m upstream.  Moving the target upstream has the effect of directing smaller-angle, higher-momentum particles into the focusing horns, resulting in a higher-energy neutrino beam, as shown in Figure~\ref{fig:numi-spectra}.\footnote{For maximal efficiency of the ME and HE beams, both the target and the downstream horn are moved with respect to the fixed first horn \cite{numitdr}.  Because of the complexity of moving horn 2, the MINOS experiment will make use only of the target positioning system, which can be accomplished {\it in situ} \cite{flexybeam}.  Such are referred to as the pseudo-Medium (pME) and pseudo-High (pHE) beams.}

\begin{sidewaysfigure}[p]
  \centering
  \includegraphics[width=9 in]{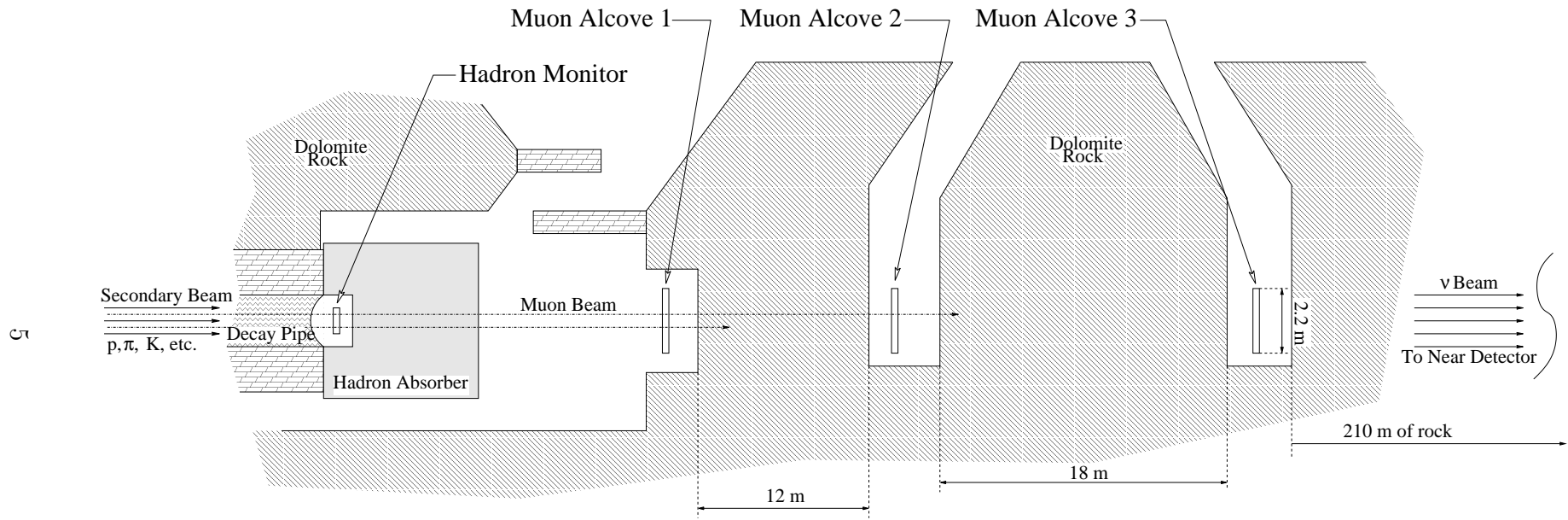}
  \caption{Plan view of the downstream areas of the NuMI beamline.
    The beam, consisting of hadrons, muons, neutrinos, and remnant protons, enters the area through the decay pipe.  The hadron
    beam's spatial distribution is measured at the Hadron Monitor and then
    stopped in the Hadron Absorber.  The higher-energy muons penetrate the
    absorber and some distance of rock; along the way their spatial 
    distributions are measured by the Muon Monitors.  }
  \label{beamon_layout}
\end{sidewaysfigure}

The subject of this paper is the secondary and tertiary beam monitoring system, located at the downstream end of the decay volume, shown in Figure~\ref{beamon_layout}.  Its purpose is to monitor the integrity of the NuMI target and of the horns which focus the secondary meson beam.  This monitoring is accomplished by measuring the intensity and lateral profile of the remnant hadron beam and of the tertiary muon beam.  Because every muon is produced by the same meson decays which produce neutrinos, the muon beam provides a good measure of the focusing quality of the horns. The large fluxes of the hadrons and muons permit the secondary beam monitors to diagnose, in a single spill, problems in the upstream neutrino beam systems.

The first accelerator neutrino beam at BNL \cite{Danby:1962nd} did not explicitly use muon beam monitors, although in subsequent runs emulsion detectors were placed in slots in the steel shielding in front of the neutrino detector to analyze the muon spectra, and thus provide a check on the muon-neutrino fluxes \cite{Burns:1965dn}. The 1965 CERN neutrino beam 
utilized a spectrometer downstream of its target station to measure secondary particle production {\it in situ} \cite{Plass:1965ds}, although such spectrometer measurements were invasive and could not be performed during normal neutrino running.  The first system which provided continuous muon beam monitoring was built at CERN for the 1967 run \cite{Pattison,Bloess}, and consisted of arrays of ionization chambers and scintillators at various depths in a muon filter.  The muon rates were used to verify the horn focusing and the neutrino flux to the experiments \cite{Bloess,Wachsmuth}. Such a continuous muon beam monitor has subsequently been used at the CERN West Area Neutrino Facility (WANF) beam \cite{Astier}, the Institute of High Energy Physics-Serpukov (IHEP) neutrino beam \cite{Anikeev,Bugorsky}, the BNL neutrino beam \cite{Chi:1989si}, and the K2K beam in Japan \cite{Hill:2001pu,Oyama:2001bp}.  

Section~\ref{overview} describes properties of the secondary hadron and tertiary muon beams at NuMI.  Section~\ref{beamon_chamber} discusses the design of the secondary beam monitoring chambers.  Section~\ref{calib} discusses the calibrations of the detectors within the arrays, necessary to ensure proper measurements of the lateral profiles of the beams.  Section~\ref{perf} discusses the chambers' performance in the high fluences of the beam.  Section~\ref{beam-meas} discusses some of the diagnostic capabilities of the beam monitors as demonstrated in the first few months of beam operation.

\section{System Overview}
\label{overview}

The Hadron Monitor, located at the end of the decay pipe, measures the rate and profile of the uninteracted beam.  Such measurements are sensitive to changes in the target and baffle system.  The flux at the center of the Hadron Monitor is dominated by protons passing through the target, so is relatively insensitive to the details
of horn focusing.  It is, however, quite sensitive to the target's geometry, as manifested by the observed attenuation of the proton beam and by multiple Coulomb scattering of the beam through the target material.  In this way, it is used on a spill-to-spill basis to ensure that no failure of the target has occurred.
Experience from BNL\cite{Mann:1993ni} and our own experience operating NuMI (see Section~\ref{beam-meas}) have shown that having an {\it in situ} monitor of the target's integrity can be quite useful.  The Hadron Monitor was also used extensively as part of a beam-based alignment check of the target and horn components, as will be discussed in a forthcoming article \cite{bba}.  Prior to installation of the target in the beam line, the Hadron Monitor was used as a check of the alignment of the proton beam transport.

Past experience from other experiments shows that it is also desirable to have measurements from Muon Monitors to diagnose problems such as component failures\cite{Mann:1993ni}, misalignments \cite{Casagrande}, or non-ideal horn optics \cite{Dusseux:1972cm}.  In contrast to neutrino detectors, in which it may be necessary to accumulate data for days or weeks in order to reveal problems, Muon Monitors can provide accurate measurements in a single beam pulse.  By sampling the muon distribution at several locations in the shielding the monitors achieve sensitivity to the energy spectrum of the muon beam.  

To reach the muon detectors in alcoves 1, 2, and 3, muons must have an initial momentum of 5, 12, and 24 GeV, respectively.  These thresholds are dictated by 6~m of shielding in the upstream beam absorber and 12~m of unexcavated rock between alcoves 1 and 2 and 18~m of rock between alcoves 2 and 3.  These muon momentum thresholds correspond to pion decays producing 3.8, 9, and 18~GeV neutrinos.  Referring to Figure~\ref{fig:numi-spectra}, which shows the neutrinos fluxes for the low, medium, and high neutrino energy configurations of the beam line calculated for the detector hall located on the NuMI beam axis, 1040~m away from the NuMI target, it is apparent that the alcoves cover an increasing portion of the spectrum for the higher energy beam configurations.  Thus, the Muon Monitors' measurement ability is enhanced in the pME and pHE beams.  While the 5~GeV muon momentum threshold limits the monitoring capability in the LE beam, this threshold is dictated by the shielding thickness in the absorber necessary to contain the hadronic shower from the remnant proton beam.\footnote{Muon monitors placed farther upstream, in the shielding, can become dominated by the remains of the hadronic shower, as has been shown previously in other experiments \cite{Zimmerman,Auchincloss}.}  Although MINOS will run primarily in the LE configuration of NuMI, periodic runs in the pME and pHE configurations are envisioned for diagnostic purposes, and future experiments may want these higher neutrino energy configurations.


\begin{table}[tp]
  \centering
  \begin{tabular}{p{4cm}|c|c|c|c|c|c}

 & \multicolumn{3}{|c|}{Fluence }&\multicolumn{3}{|c}{Beam RMS }\\

 & \multicolumn{3}{|c|}{(10$^5$/cm$^2$/10$^{12}$ppp)}&\multicolumn{3}{|c}{ Size (cm)}\\

    & {LE} & {pME} & {pHE}& {LE} & {pME} & {pHE} \\
    \hline
    \centering { Hadron Monitor} & 
     680 & 680 & 680 & 20. & 20. & 20.  \\ \hline
    \centering {Muon Alcove 1} & 6.5 & 10.0 & 9.0  & 190 & 130 & 110\\ \hline
    \centering {Muon Alcove 2} & 0.9 & 5.0 & 7.2  & 250 & 140 & 110  \\ \hline
    \centering {Muon Alcove 3} & 0.35 & 0.5 & 2.3 & 190 & 250 & 120 \\
  \end{tabular}
  \caption{Predicted maximum particle fluence and beam size at the
    monitoring stations for different types of beam. 
    {LE} is the low-energy beam configuration to be used extensively for the MINOS experiment, {ME} and {HE} are the      
     medium and high energy beam configurations described in \cite{flexybeam}.  The beam size is the FWHM of the lateral distribution of particles in the beam at a given monitor station}
  \label{tab:beamon_rates}
\end{table}

Table~\ref{tab:beamon_rates} shows the calculated peak charged particle fluence at each detector for each of the NuMI neutrino beam configurations.  The fluence at the hadron monitor is estimated by our beam Monte Carlo to be comprised of over 60\% protons within the first 0.5~m in radius from the beam axis, with the rest coming from pions and electromagnetic showers.  The proton beam size at the hadron monitor is estimated to be $\approx$20~cm, due to multiple scattering of these protons through the 0.94~m graphite target 725~m upstream.

The muon beam is very broad at the Muon Monitors, its extent being set largely by pion decay kinematics and the decay pipe size.  The breadth of the muon beam allows us to merely sample the muon profile with detectors in the alcoves over the first 1~m in radius.  While the muon profile is quite broad in the LE beam (see Table~\ref{tab:beamon_rates}), the profile in the pME and pHE beams is sufficiently narrow to permit position measurements precise to $\sim1$~cm.\footnote{Such precision is better than the 2~cm survey accuracy of the muon alcoves' positions.}  Periodic runs in these higher energy beams are therefore envisioned \cite{flexybeam}.  The fluences at the Muon Monitors are almost entirely muons.  Even in the first alcove, located behind the absorber, calculations and measurements of the particle rates taken with the horns and target removed indicate that neutron background in this alcove emanating from the beam absorber is $<1$\% \cite{keisler}.

The beam Monte Carlo fluences imply a yearly dose of 2~GRad at the center of the Hadron Monitor, dominated by protons from the beam and neutrons emanating from the beam absorber.  The Muon Monitor in alcove 1 is exposed to $\sim80$~MRad/year in the LE beam, coming mostly from muons.  Both of these doses assume a yearly accumulation of $4\times10^{20}$ protons on target.

\begin{figure}[t]
  \centering
  \includegraphics[width=6.5cm]{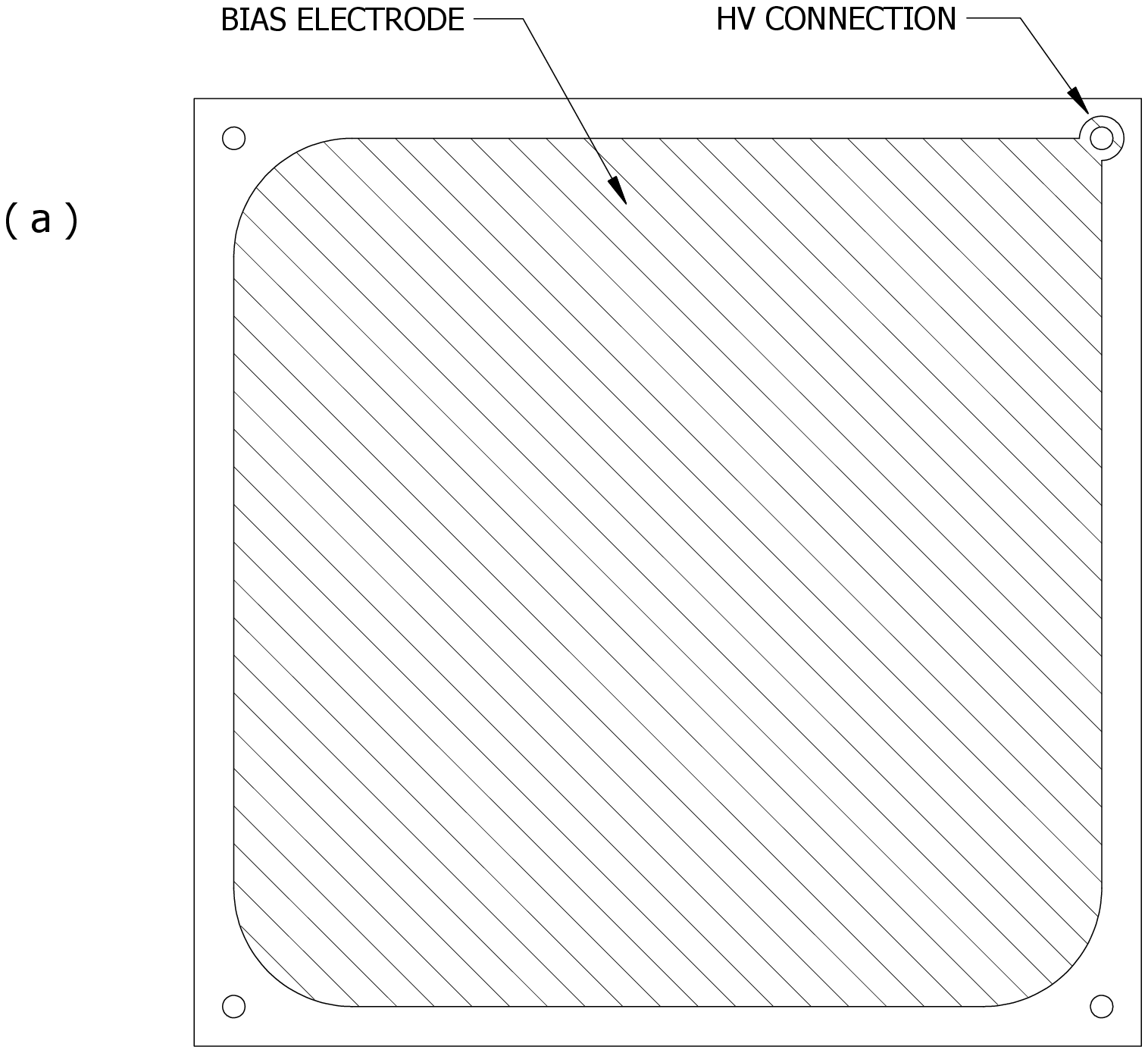}
  \includegraphics[width=6.5cm]{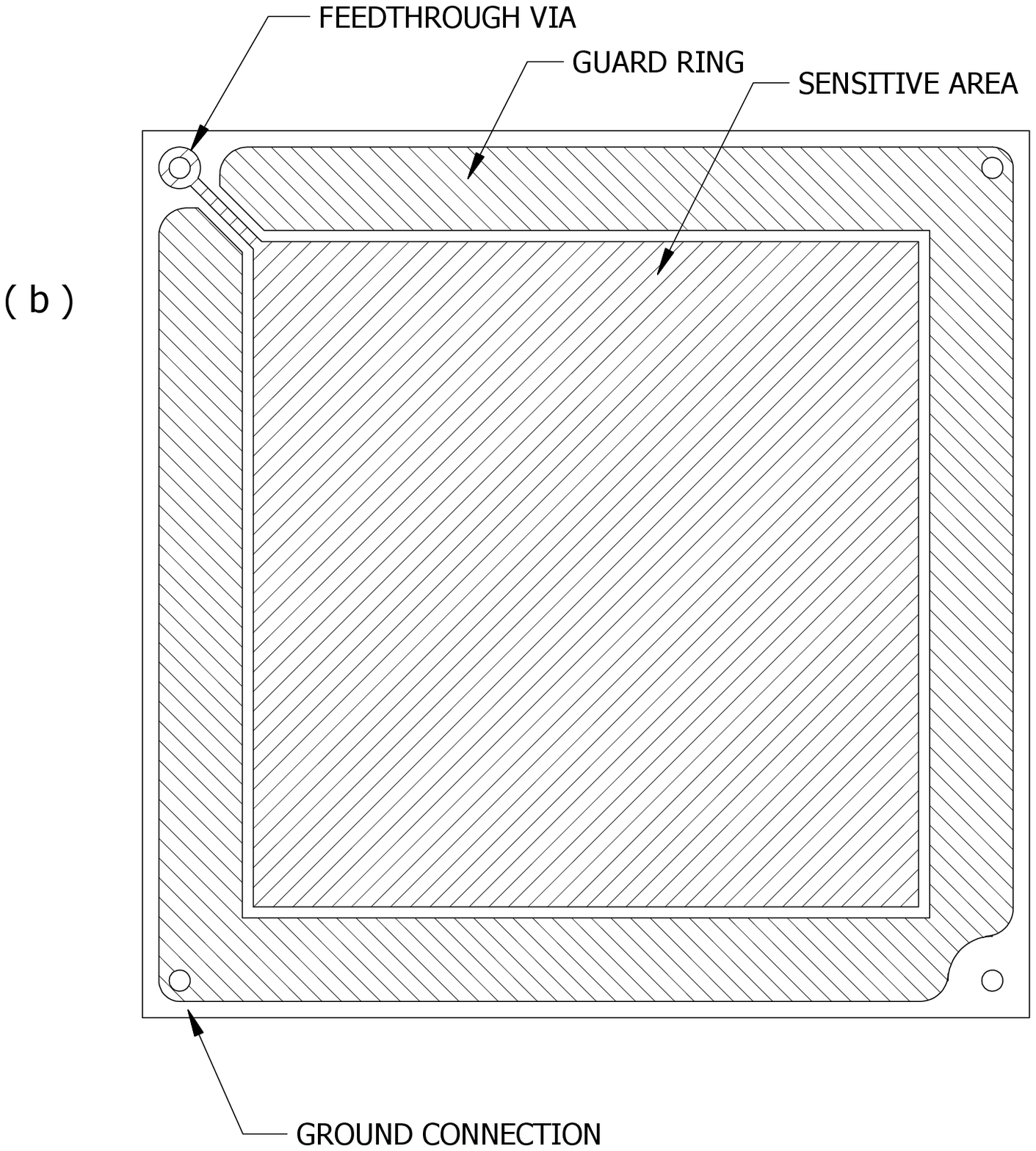}
  \caption{The ceramic plates that make up each vessel.  Both have
  silver platinum electrodes facing the the ionization volume.  
  ({\em a}) Bias voltage plate with a single electrode that connects
  to the bias supply through a corner post.  ({\em b}) Signal plate
  with sense pad and grounded guard-ring electrode. }
  \label{beamon_chamber_plates}
\end{figure}

\section{Ionization Chamber Design}
\label{beamon_chamber}

The Muon and Hadron Monitors consist of arrays of ionization chambers.  Each charged particle passing through the chamber ionizes Helium gas, with the charge drifting to the chamber electrodes being proportional to the incident particle flux. Parallel plate ionization chambers were chosen with an electrode spacing of 1~mm for the Hadron Monitor and 3~mm for the Muon Monitors.
As demonstrated by previous beam tests \cite{rdf,atf}, such electrode spacings ensure full charge collection without recombination in the chamber gas.  The parallel plates are made from ceramic wafers with Silver-Platinum electrodes, offering radiation tolerance and better mechanical precision for the smaller electrode spacings \cite{atf}.  The chamber plates in Figure~\ref{beamon_chamber_plates} are alumina ceramic cut to 4'' squares, with the separations defined by ceramic washers.  

\begin{figure}[t]
  \centering
  \includegraphics[width=12cm]{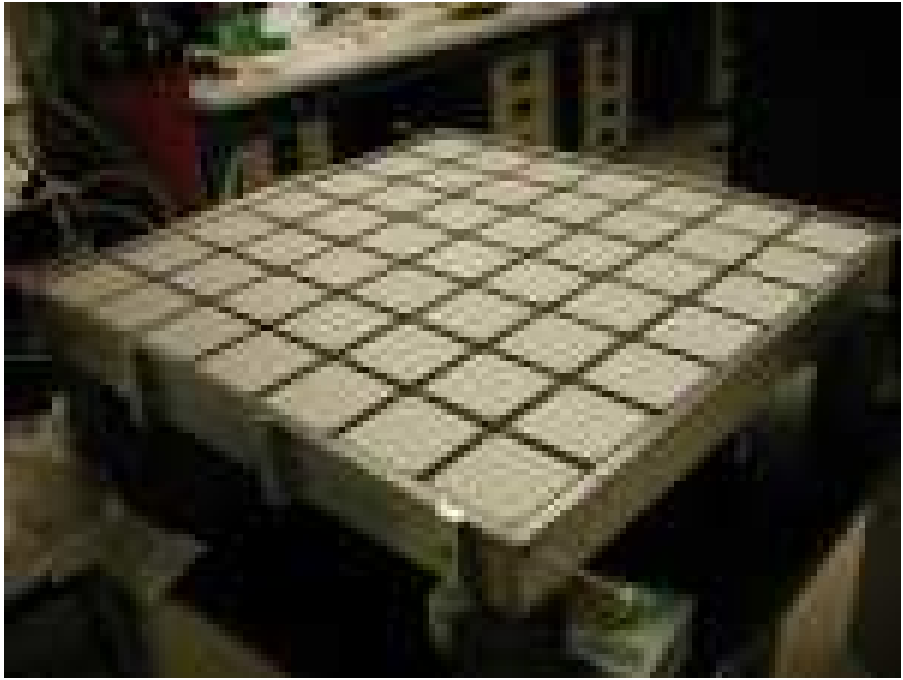}
  \vskip -.2 cm
  \caption{View of the Hadron Monitor interior.  The 49 chamber array
    measures about 1 m square.  The poor image quality is necessitated by {\tt arXiv}'s file quota.  For an acceptable image, see the Fermilab preprint server or NIM's server.}
  \label{beamon_had_iso}
\end{figure}

\subsection{Hadron Monitor}

The Hadron Monitor is a 7$\times$7 array of 1 mm gap ionization chambers in a single gas vessel.  The chambers are spaced 11.4~cm center-to-center, for a total lateral extent of 76~cm horizontally and vertically.  The vessel is an Aluminum box with a cover that is sealed to the box with a solder-wire gasket.  Each high voltage and signal channel has its own ceramic feedthrough, custom high-radiation cable, and coaxial cables back to the equipment racks outside of the radiation area;  this separation allows individual channels to be disconnected in the case of a failure.  The Hadron Monitor interior is shown in Figure~\ref{beamon_had_iso}.

\begin{figure}[t]
  \centering
  \includegraphics[width=14cm]{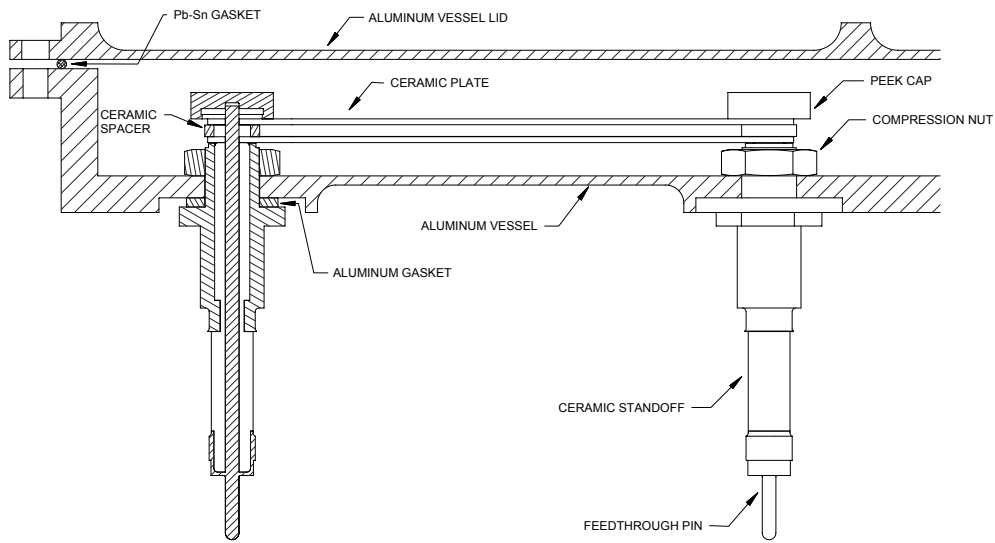}
  \vskip -.2 cm
  \caption{Profile of the Hadron Monitor feedthrough and chamber
    assembly.  The feedthrough is a modified vacuum high-voltage
    feedthrough that is gasketed to the Aluminum vessel.  The 
    chamber plates mount directly on the feedthroughs.  A PEEK cap
    is attached to the top of the feedthrough lead to avoid
    collecting stray ionization.}
  \label{beamon_had_feedpic}
\end{figure}

The Aluminum vessel is as thin as possible to minimize radioactivated mass.    A cross sectional view of one of the chambers mounted in the vessel is shown in Figure~\ref{beamon_had_feedpic}.  The feedthroughs are high-vacuum components with ceramic insulators rated to 10 kV.  The ceramic plates are supported at two corners by these feedthrough pins, and at the other two corners by Aluminum standoffs.

The Hadron Monitor is installed inside the absorber shielding and is exposed to very high radiation levels; hence, particular attention was given to its cabling.  Calculations show that the radiation levels fall off by a factor of $\sim3$ at 1~m from beam center, and another factor of 10 through the concrete shielding.  A redundant cable was designed for the span behind the Hadron Monitor, where the radiation levels are highest: a coaxial Kapton cable was stripped of its outer
ground braid for its last $\sim$ 1~m and a ceramic tube slid over for extra insulation.  The ceramic has an Aluminum sleeve around it, acting as the exterior ground; each sleeve was soldered to the kapton cable's braiding at the junction.

The Hadron Monitor sits directly in front of Hadron Absorber
and is only accessible by a 6'' $\times$ 40'' slot in the side of the
absorber shielding.  The Hadron Monitor is inserted by sliding it in on pre-installed rails. 
Its alignment is defined by the location of the rail
and the stop on that rail.  

\subsection{Muon Monitor}
\label{beamon_mu}

\begin{figure}[t]
  \centering
  \includegraphics[width=13.4cm,angle=270]{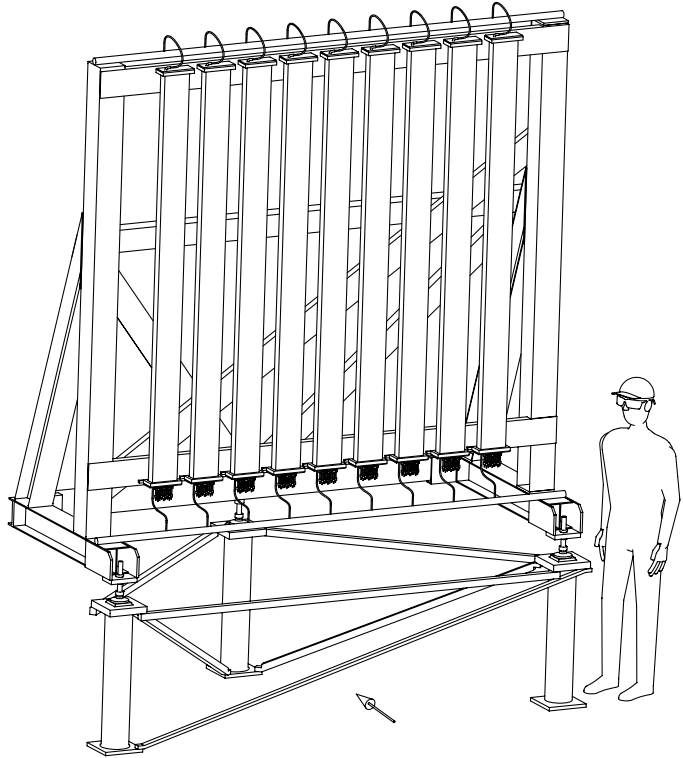}
\vskip -1cm
  \caption{One of the three Muon Monitor arrays.  There are nine 
    ionization chambers in each of nine tubes.  }
  \label{beamon_mu_array}
\end{figure}

\begin{figure}[t]
  \centering
  \vskip -.0cm
  \includegraphics[width=5.5 in ]{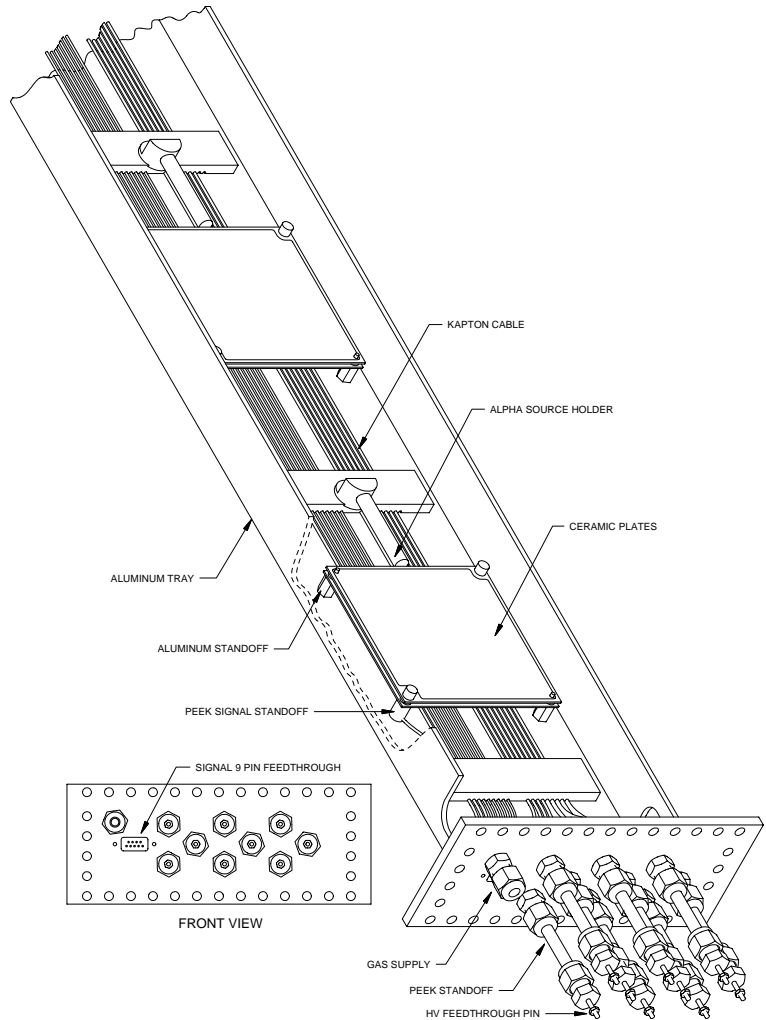}
  \vskip -2.2 cm
  \caption{View of a portion of a Muon Monitor ``tray''.  The chambers
    are mounted to an Aluminum channel and the cables run down the 
    length to the flange.}
  \label{beamon_mu_tray}
\end{figure}

The Muon Monitors are composed of three 9$\times$9 arrays of 3 mm 
ionization chambers.  The arrays are each made up of nine ``tubes'' (see Figure~\ref{beamon_mu_array}).
Each tube contains a tray with nine chambers mounted, as shown in 
Figure~\ref{beamon_mu_tray}.  The chambers within each tube
are spaced 25.4~cm center-to-center, likewise for the center-to-center distance 
between muon tubes.  Each tube is an
independent Helium volume with electrical feedthroughs and gas 
connections.

Each of the nine chambers within one tray is mounted to the tray using four standoffs.  Kapton-insulated coaxial cables are routed on the tray toward one end.  A 1~$\mu$Ci Americium-241 source, intended as a calibration signal, is mounted next to each ion chamber.  The high-voltage cables are routed into custom feedthroughs made with PEEK insulators and compression fittings, while the signal cables were all soldered onto the pins of a 9-pin D-type ceramic vacuum feedthrough.  A PEEK collar fit over the cables and pins to insulate the conductors from the gas in order to prevent stray ions from collecting on the conductors \cite{zwaska}.

\subsection{Gas System}
\label{beamon_gas}

The beam monitors operate with a continuous flow of Helium gas supplied by a manifold of ultra-high purity $(\sim1-10$~p.p.m. contaminant) cylinders.  A line pressurized to 4~atm. delivers the Helium to a distribution rack 400~m away, down in the beam line tunnel.  The distribution rack splits the line into four, one for each ion chamber array, and establishes each array's flow.  The flow and pressure are measured on the lines to the detector arrays.  The precision of the pressure transducers is $\sim$0.1~Torr and of the flowmeters is $\sim$0.1~$\ell$/hr.  Each line is fitted with a pressure relief valve: the Hadron Monitor is limited to 15~Torr overpressure and the Muon Monitors to 150~Torr overpressure.  

The pressure and temperature at the detectors was monitored by a second set of pressure transducers and by resistive thermal devices (RTD's).  The pressure at the distribution rack and at the detector were observed to track each other within 0.2~Torr.  Within one month of operation of the beam, radiation damage to the pressure transducers in the beam enclosure made the pressure transducers there inoperable.  We subsequently relied on the redundant tranducers at the distribution rack.  

At a typical flow of 25~$\ell$/hr. to the Hadron Monitor, measurements indicate that it achieves an impurity level of 80~p.p.m.  A flow of 10~$\ell$/hr. to each of the muon alcoves was sufficient to achieve 20~p.p.m. impurities in those detectors.  The higher impurity level in the Hadron Monitor is due to a leak which developed during its assembly.  These flow rates correspond to 30 and 1.3 volume exchanges per day for the Hadron and Muon Monitors, respectively.  


\section{Chamber Calibrations}
\label{calib}
In order to provide 50~$\mu$rad pointing precision of the proton beam, a 5\% relative chamber-to-chamber calibration within the Hadron Monitor is required (irrespective of NuMI beam configuration).  Monte Carlo studies indicate that such a 5\% relative calibration of each chamber provides a proton beam centroid determination to within 3cm, consistent with the limits of the optical survey of the location of this detector relative to the NuMI target which is 725~m upstream.  Such an error corresponds to a 42 $\mu$rad uncertainty in the proton beam direction.

In order to perform diagnostics of the low-energy beam configuration of NuMI, the Muon Monitor requires a 1\% relative calibration of its ion chambers in order to achieve a beam alignment of 100 $\mu$rads and to permit use of the relative pulse heights in each alcove as a coarse check of the beam's energy spectrum.  Further, Monte Carlo studies indicate that variations in the horn current or mechanical alignment relative to the target will result in changes in the muon beam intensity and profile at a level requiring 1\% sensitivity to detect these effects at the $3\sigma$ level.

The relative calibration of each chamber in the muon and hadron monitors was achieved by mapping each chamber with a 1~Ci Am$^{241}$ gamma source \cite{dharma06}.  The induced ionization current was compared from chamber to chamber after systematic effects from temporal variations from electronics drift, gas pressure and temperature were removed.  

Three features of the calibration apparatus aided in the control/monitoring of systematic variations over the 1~year period in which calibrations were conducted.  First, the gas system purged the chambers with pure gas and had instrumentation for measuring pressure, temperature, and impurity levels; the knowledge of the pressure change, along with the measured change in response of the ion chambers with pressure and temperature, permitted equilibration of all the calibration constants derived over the course of the year.  Second, the electronics were re-calibrated for drift with each Hadron or Muon chamber to be calibrated.  Third, a control/reference ion chamber, with its own internal calibration source, was mounted in series with the chambers being calibrated in the gas system;  any temporal variations in the gas system would thus be observed in the control, or reference, chamber.  
The hadron monitor could be tested within 8 hours, over which time systematic drifts or pressure changes in the gas were not significant.  The 32 Muon Monitor tubes, having been constructed over the period of approximately one year from September, 2003, to August, 2004, were each calibrated in less than 8 hours, but the time between calibrations of consecutive tubes could be up to weeks.

\begin{figure}[t]
\begin{center}
\includegraphics[scale=.47]{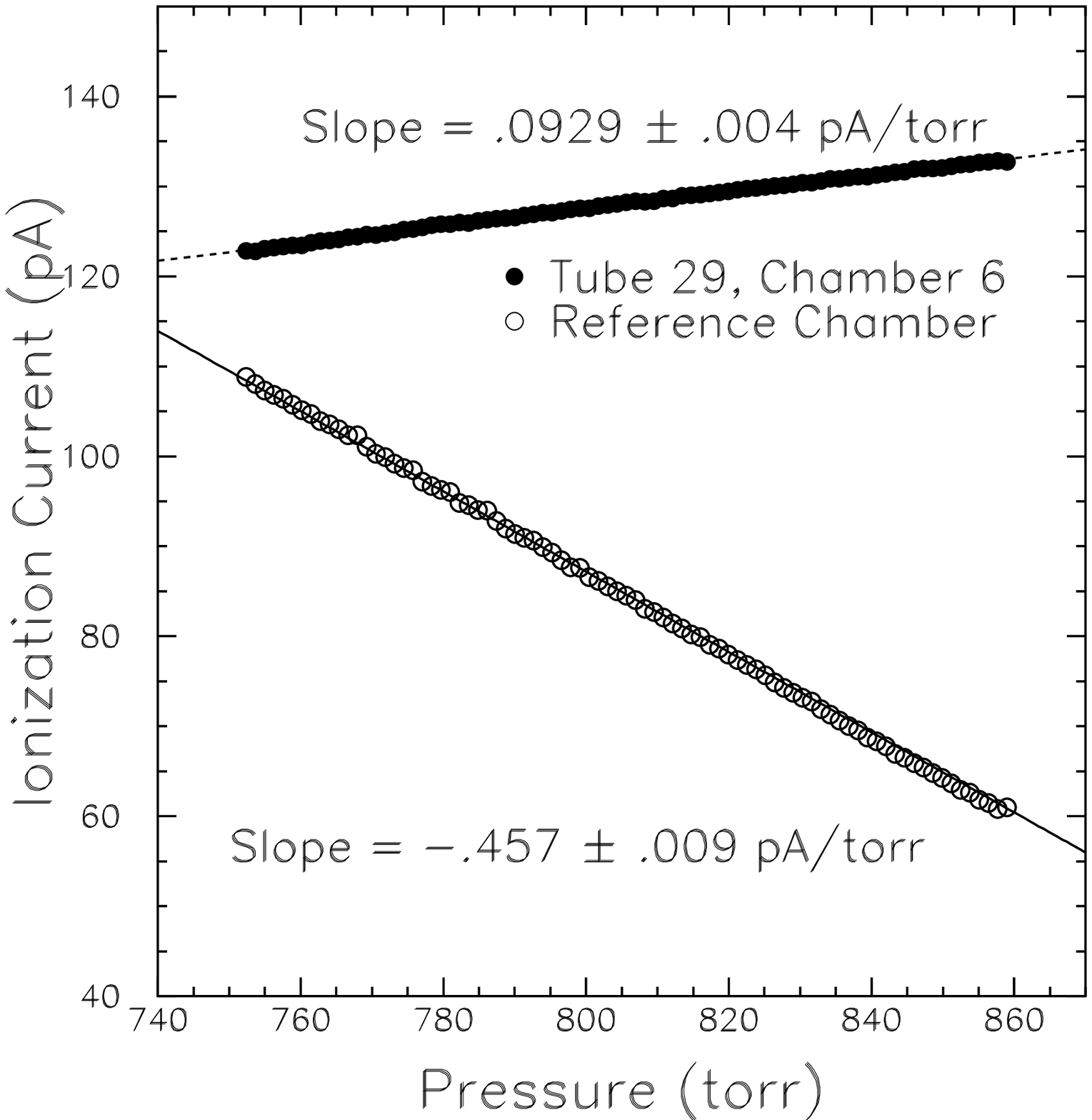}
\includegraphics[scale=.47]{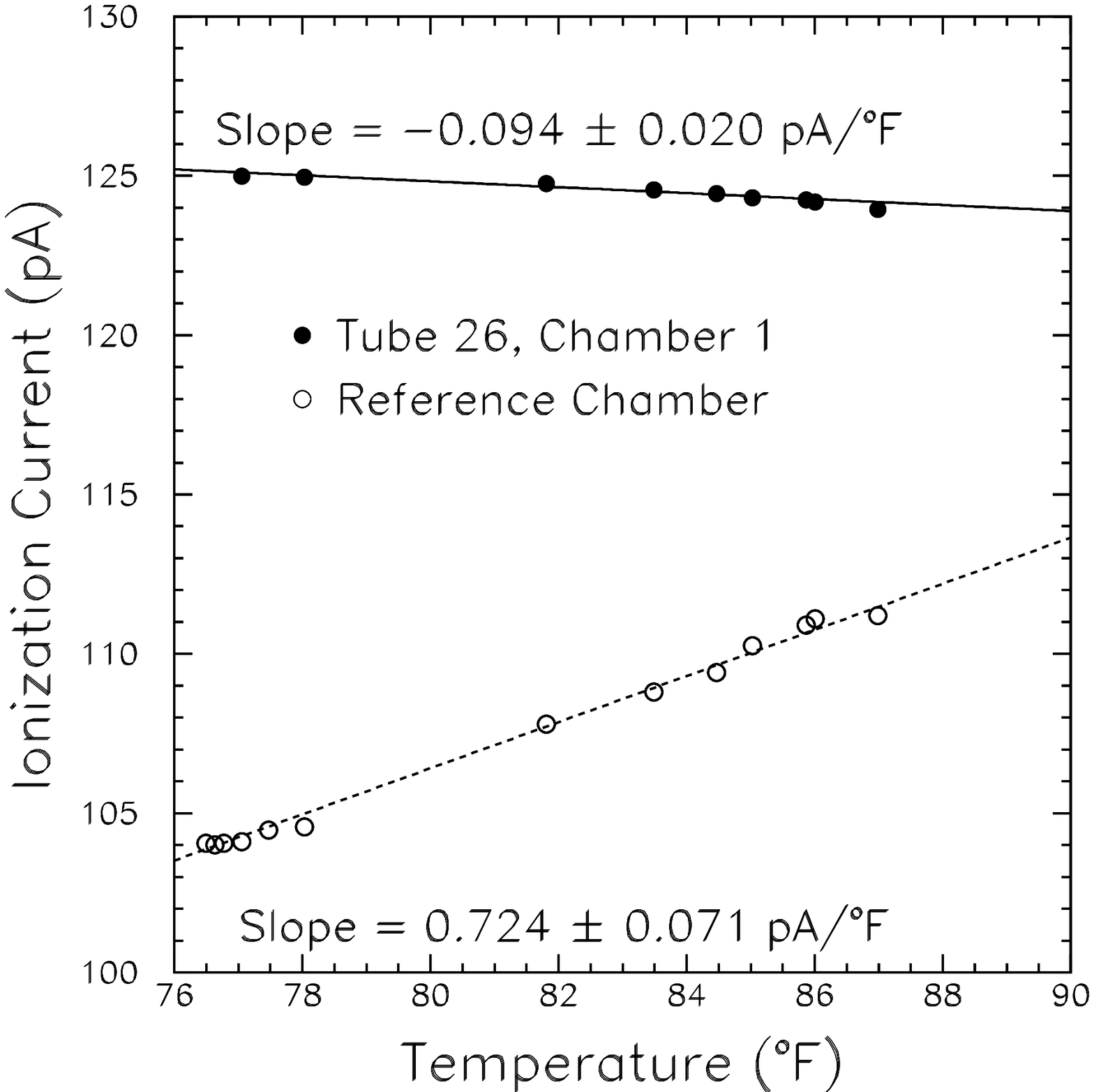}
\caption{Change in response of the ionization current in variation with pressure (top) and temperature (bottom).  Both the reference chamber and Chamber 6 of Muon Tube 26 were studied over a several hour period in which each of these two parameters was varied separately.
\label{fig:presscan}}
\end{center}
\end{figure}

Deriving calibration constants for the ion chambers required knowledge of their variation in response with temperature and pressure.  Figure~\ref{fig:presscan} shows how the chambers' response varies if either pressure or temperature is varied, derived from several hour bench studies.  The slopes in these graphs serve as our correction factors over time.  The slopes for the reference chamber are deliberately larger than for the muon tubes so as to be more sensitive to environmental changes which could affect the Muon Monitor calibrations.  Note that the reference chamber shows the opposite change in response as do the chambers being calibrated.  The ion current in the reference chamber results from a 40~$\mu$Ci Am$^{241}$ alpha source, and increasing gas density causes alphas to range out before entering the reference chamber's active volume.  Such is to be contrasted with the gamma source used to calibrate the chambers or the muons in the NuMI beam.

\begin{figure}
\begin{center}
\includegraphics[scale=.7]{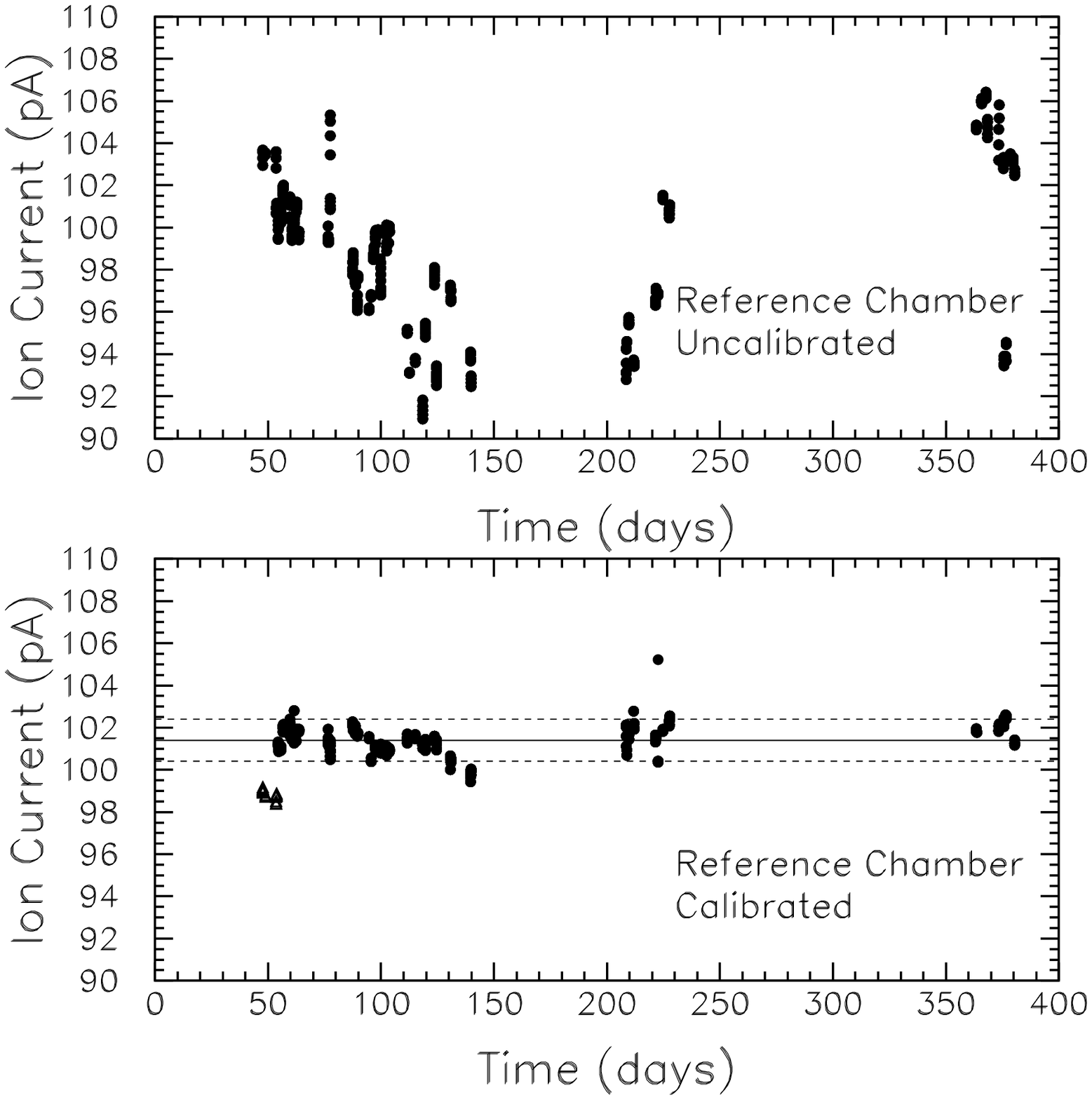}
\caption{The ionization current in the reference chamber over the 400 days of calibration operations, plotted without (above) and with (below) the corrections for pressure and temperature variations.
\label{fig:refcalvstime}}
\end{center}
\end{figure}

As a demonstration that these temperature and pressure corrections perform properly over long time periods, we studied the 
ionization current from the reference over the course of the 400~days of calibrations.  During this time the barometric pressure varied by as much as 20~Torr and the ambient temperature by 20$^\circ$F.  Figure~\ref{fig:refcalvstime} shows the ionization current measurements of the reference chamber obtained during the 400~days of testing, before and after applying the temperature and pressure corrections derived from Figure~\ref{fig:presscan}.  The calibrations leave a RMS spread of only 0.6~pA out of a signal of 101.4~pA.  Thus the reference chamber can be calibrated to better than 1\%.  Furthermore, because the reference chamber is five times more sensitive to pressure and temperature variations than the actual Muon and Hadron Monitor ion chambers (c.f. Figure~\ref{fig:presscan}), we conclude that the gas monitoring system satisfactorily controls for such variations.


The Hadron Monitor was calibrated several times to check for consistency.  A series of three full calibrations were performed in which all 49 chambers were tested.  Additionally, two partial calibrations were performed to repeat measurements on one or two chamber rows only.  The ionization currents from every chamber have been divided by the current from chamber~25 (the middle chamber) within a given run.  This scaling corrects for pressure or temperature changes that occurred in between calibration runs (again, assuming that the pressure change within a calibration run was small).  The variations of each chamber from calibration run to calibration run are within 5\%, and have an RMS of 1.2\%.

It was impractical to repeat the muon chamber calibrations, as was done for the hadron monitor.  However, we did test one tube at five different conditions spread out in time over the duration of the 400~days.  Two conclusions are drawn: 1) the multiple tests provide a 1\% agreement for any given chamber, and 2) the chambers all scale the same way with pressure and temperature, so that the relative calibration maintains its integrity.

\section{Ionization Chamber Performance in the NuMI Beam}
\label{perf}

\subsection{Measured Particle Distributions}
\label{perf_profs}

The 4 ionization chamber arrays provide two-dimensional transverse profiles of the hadron and muon beams.  During the early commissioning of the NuMI beam line, the proton beam direction and neutrino beam configurations were deliberately varied so as to benchmark the response of the secondary beam monitors to these variations.  Here, we show a few demonstrative distributions and profiles for different beam configurations.

\begin{figure}[p]
  \centering
  \includegraphics[width=7.cm]{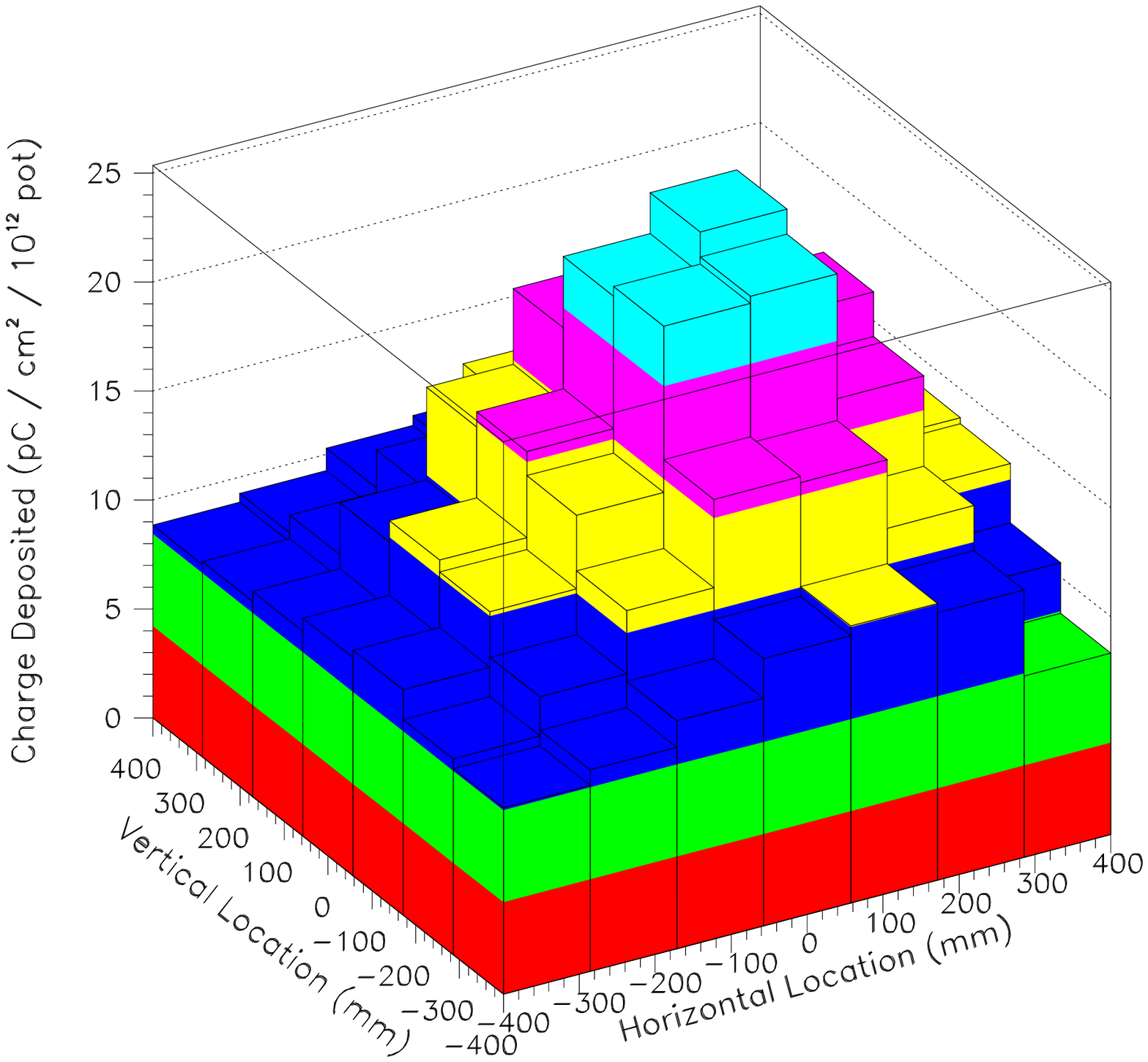}
  \hskip -.6cm
  \includegraphics[width=7.cm]{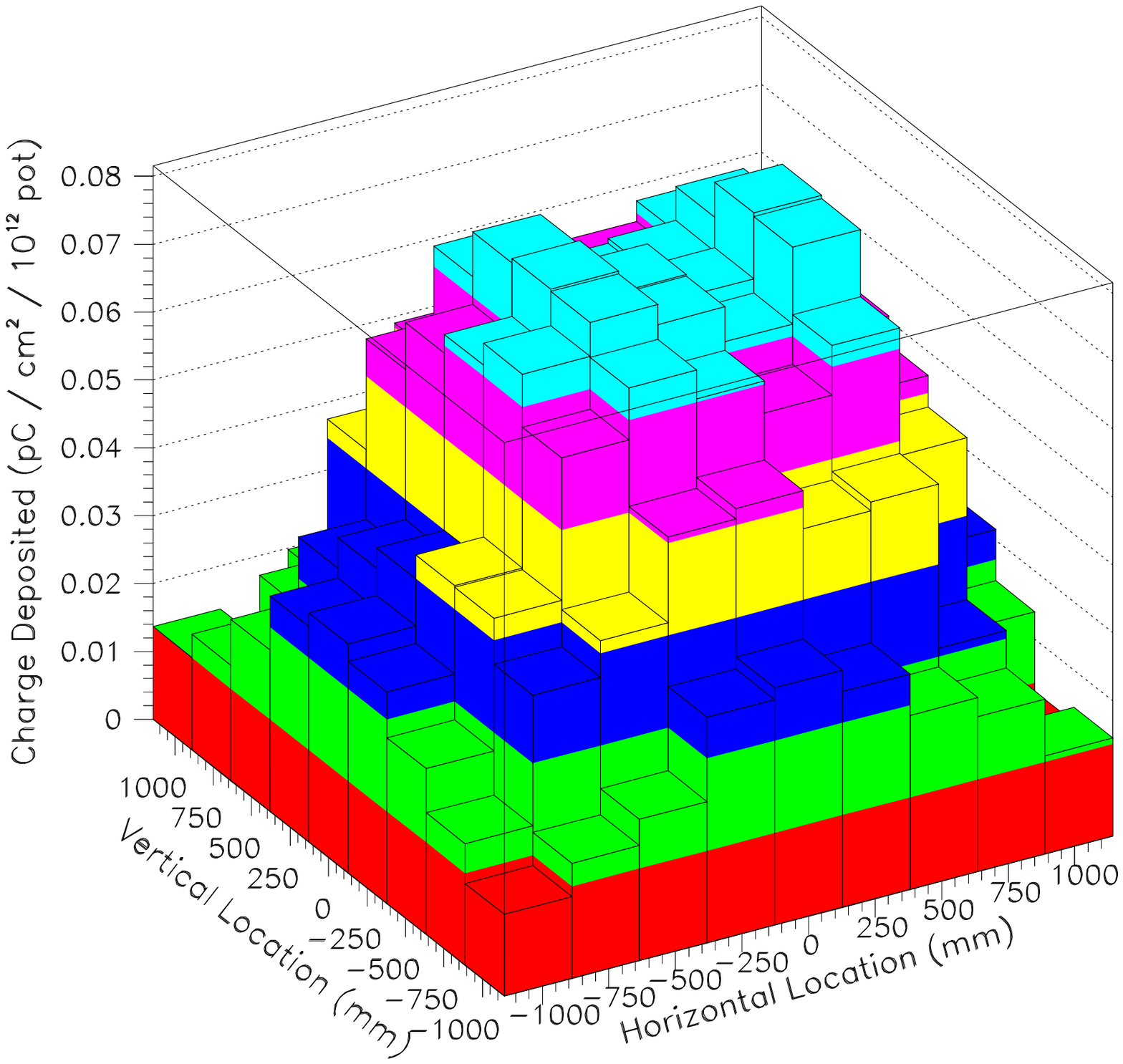}
  \vskip -.6cm
  \includegraphics[width=7cm]{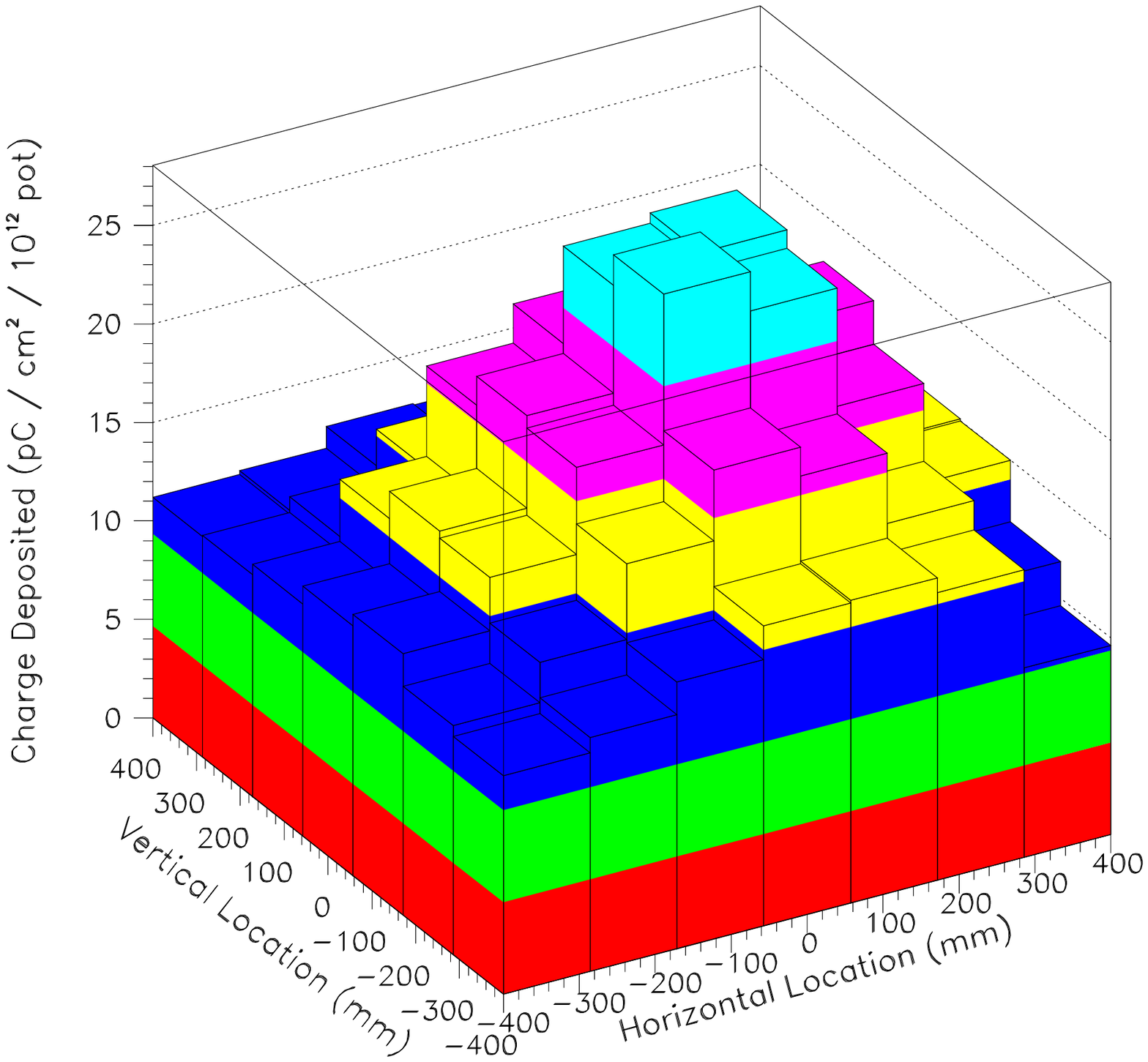}
  \hskip -.6cm
  \includegraphics[width=7cm]{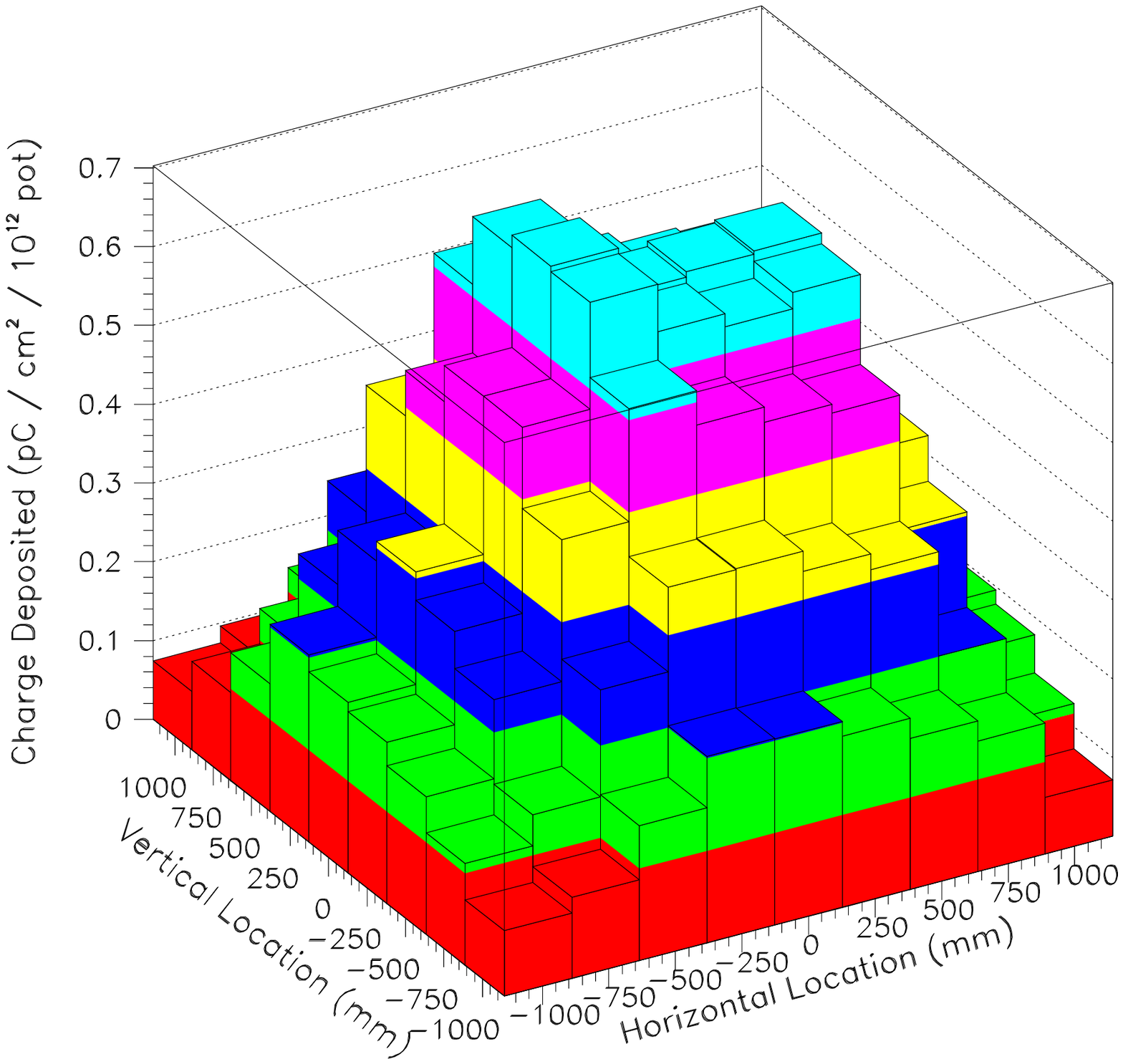}
  \vskip -.6cm
  \includegraphics[width=7cm]{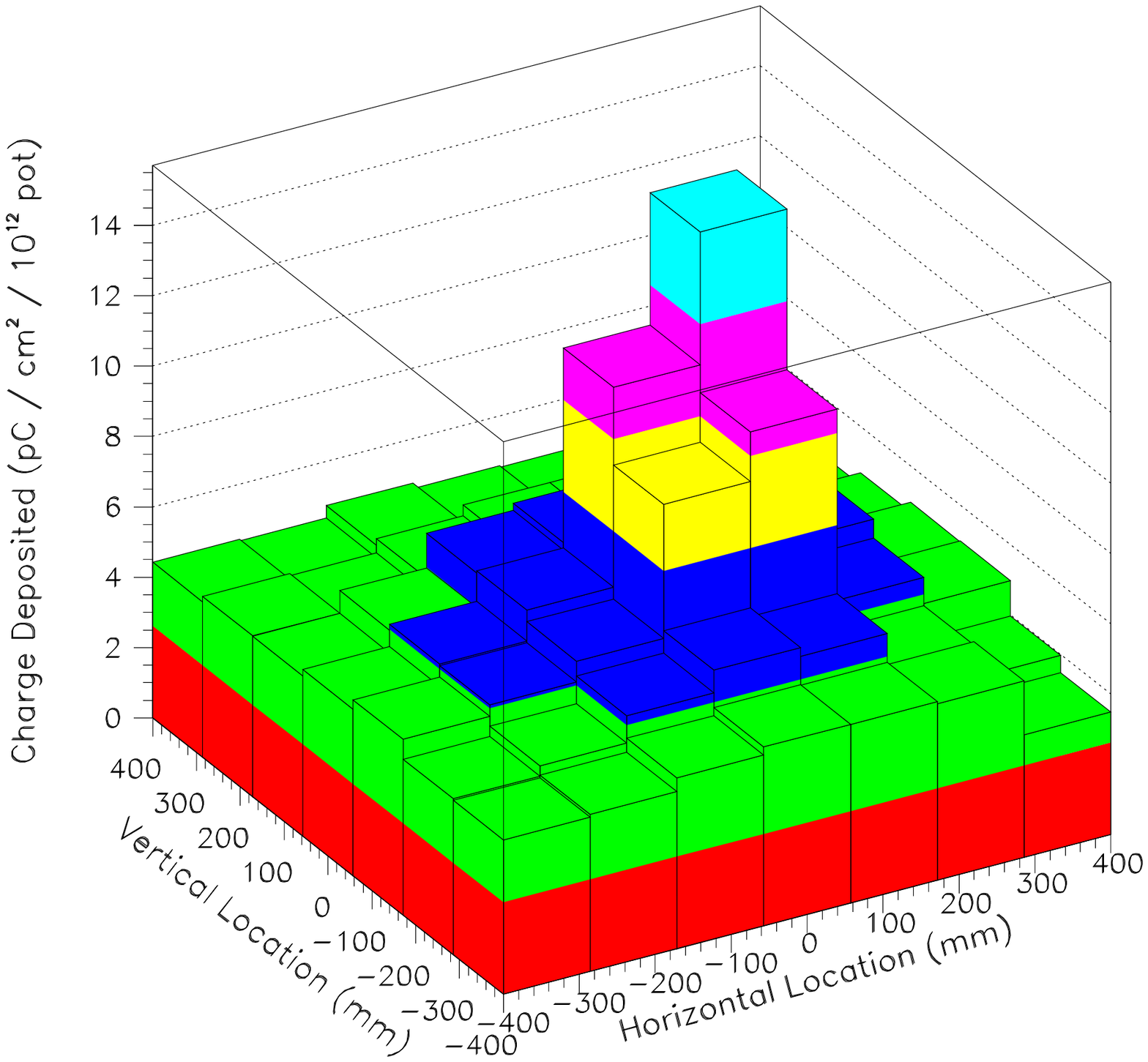}
  \hskip -.6cm
  \includegraphics[width=7cm]{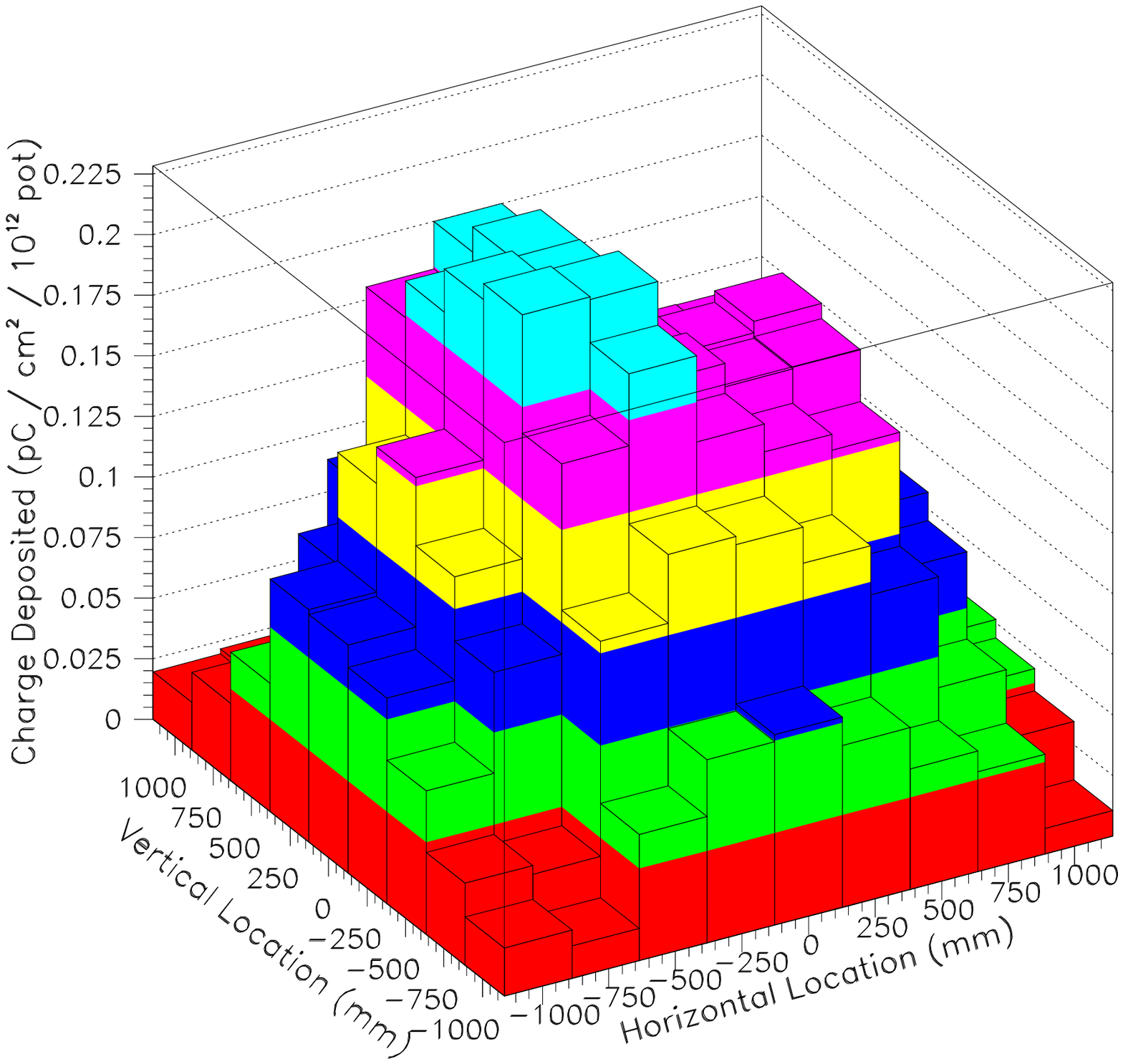}
  \vskip -.6cm
  \caption{Measured charge distributions at the hadron monitor (left column) and first muon
  alcove (right column).  The upper row shows a single beam pulse delivered to the target
  in which the horns were turned off.  The middle row shows a beam pulse in which the target
  is located in the ME position and the horns are turned on.  The lower row shows a single 
  beam pulse in which the beam, with the target in the ME position, is mis-steered into the
  upstream baffle.}
  \label{fig:perf-had-mu1}
\end{figure}

\begin{figure}[p]
  \centering
  \includegraphics[width=7.cm]{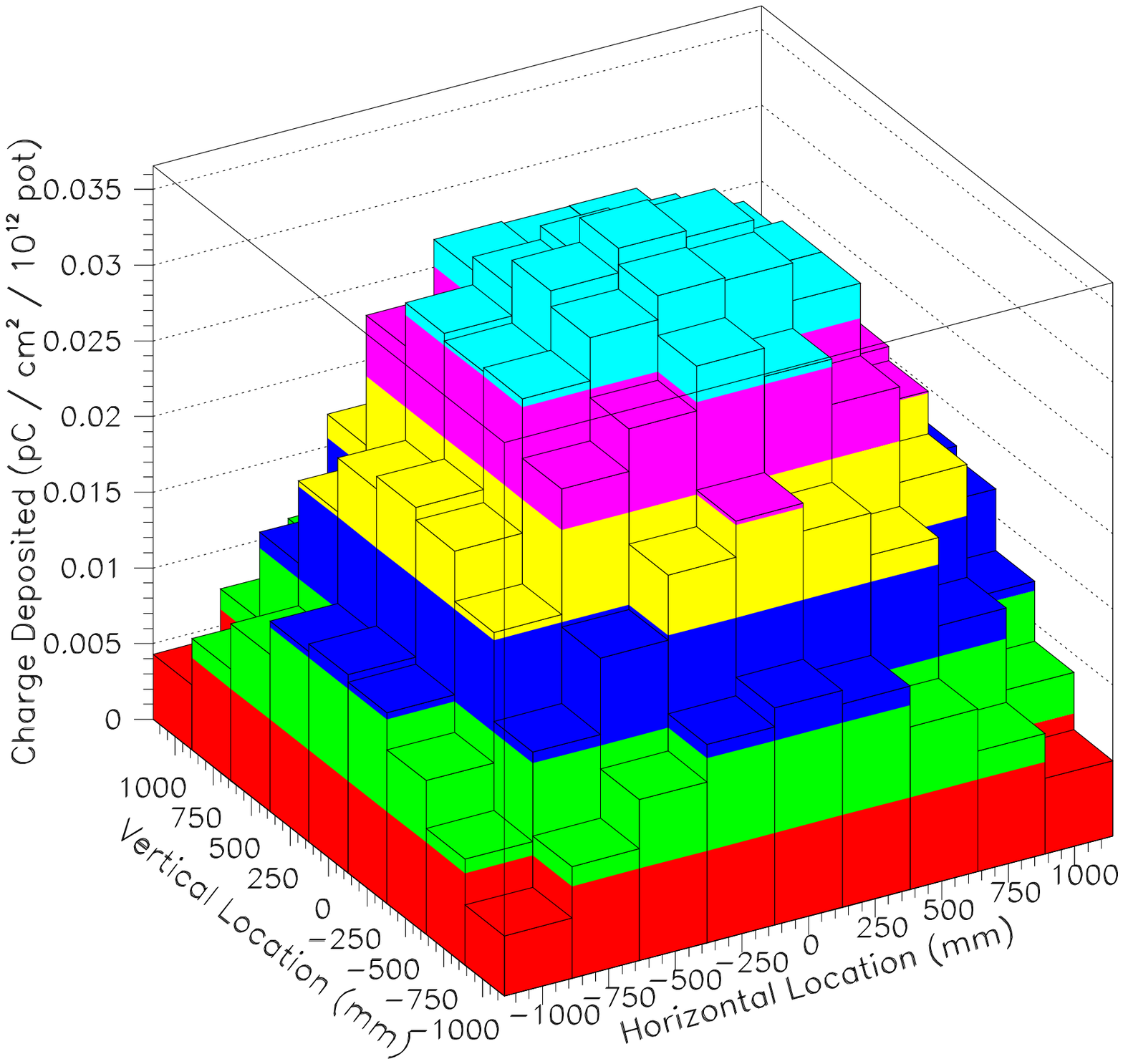}
  \hskip -.6cm
  \includegraphics[width=7.cm]{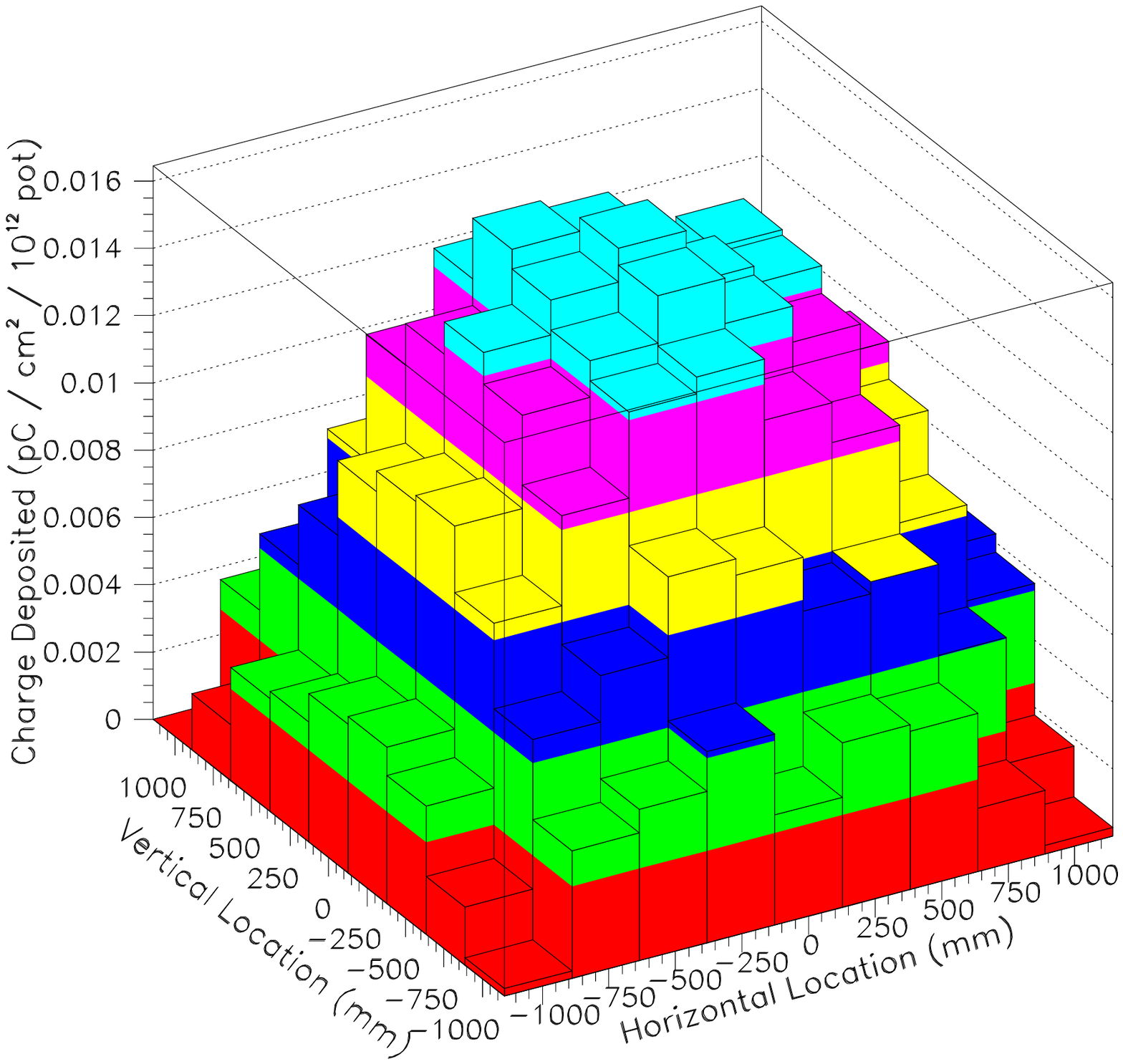}
  \vskip -.6cm
  \includegraphics[width=7cm]{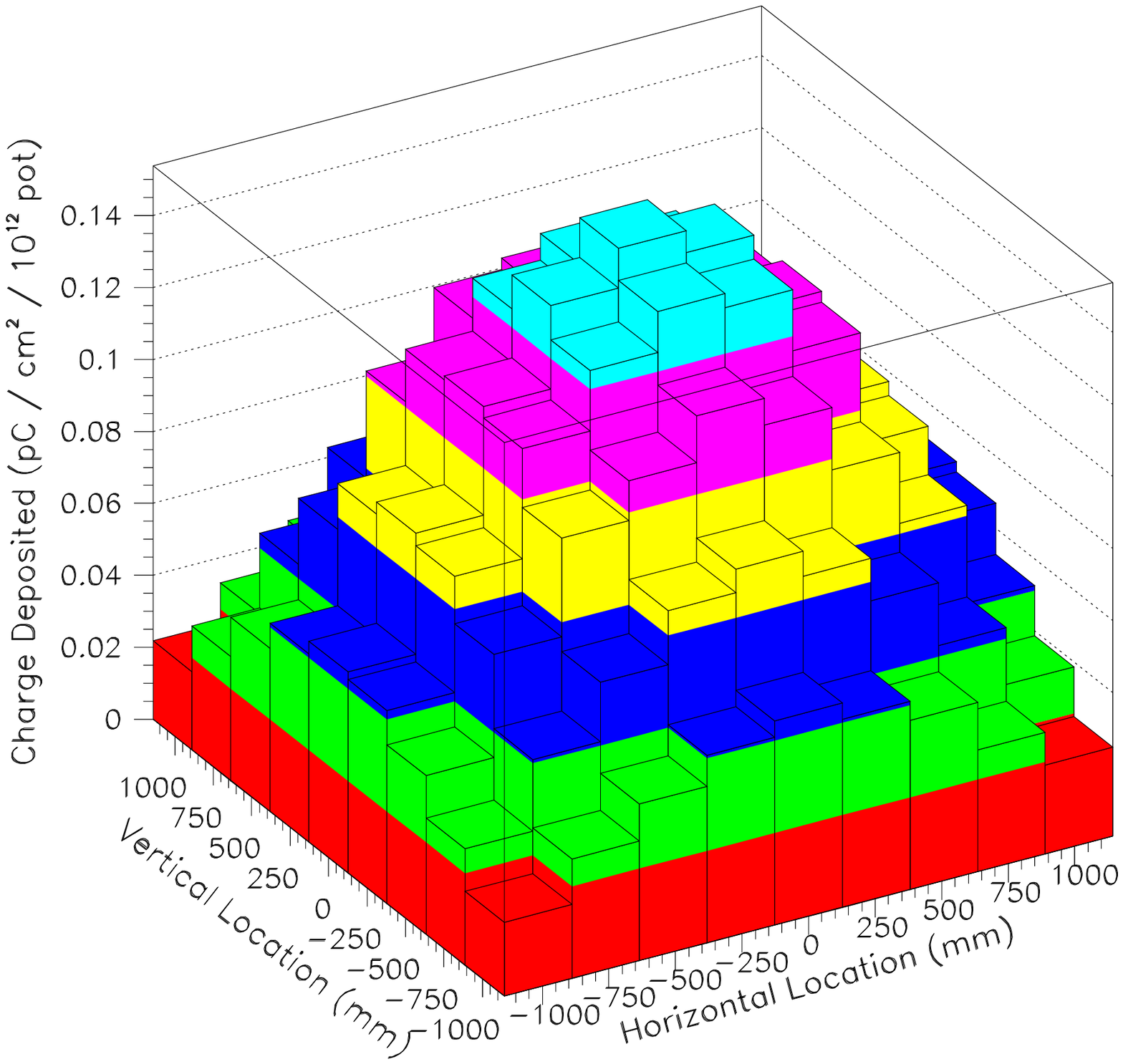}
  \hskip -.6cm
  \includegraphics[width=7cm]{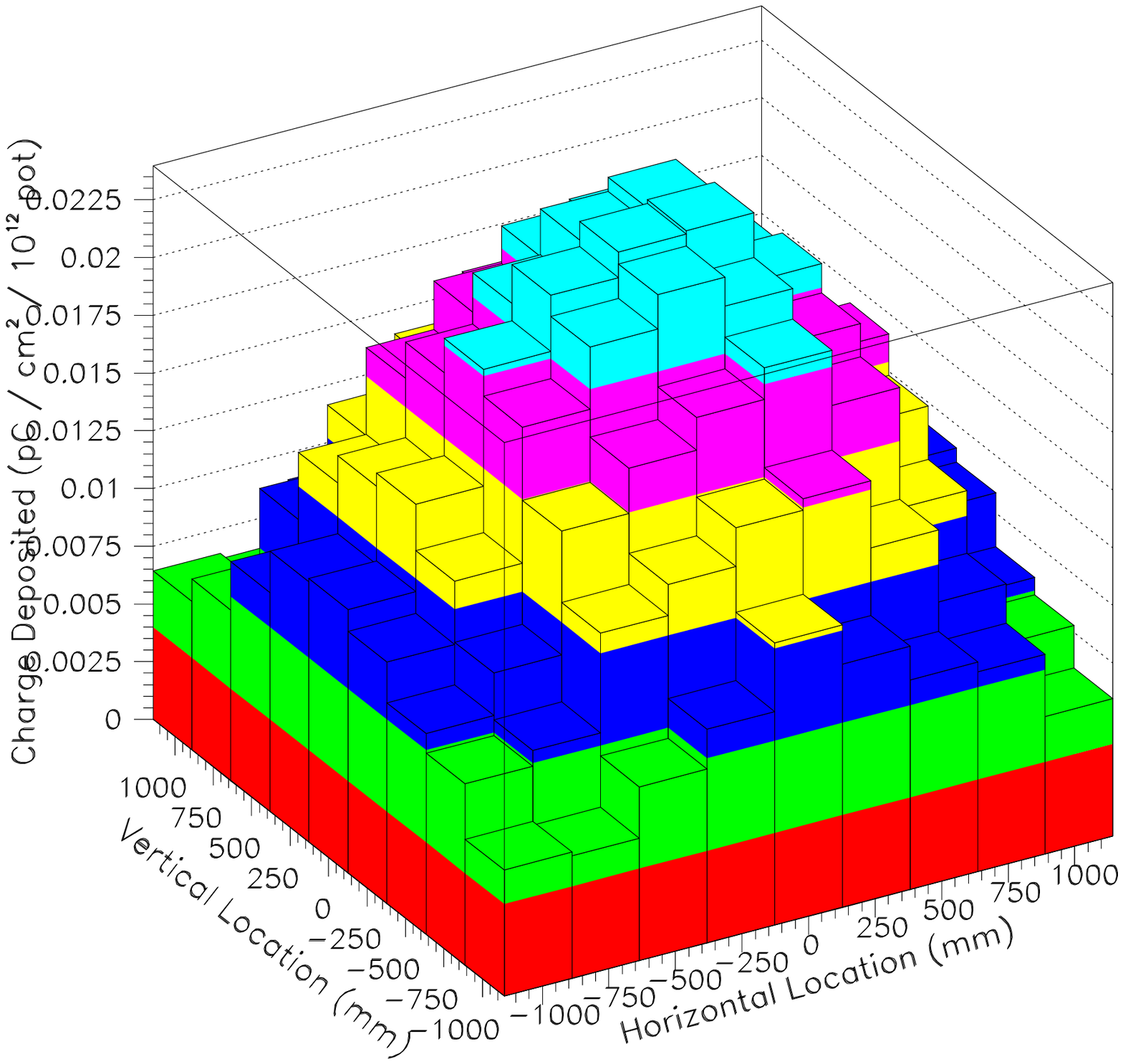}
  \vskip -.6cm
  \includegraphics[width=7cm]{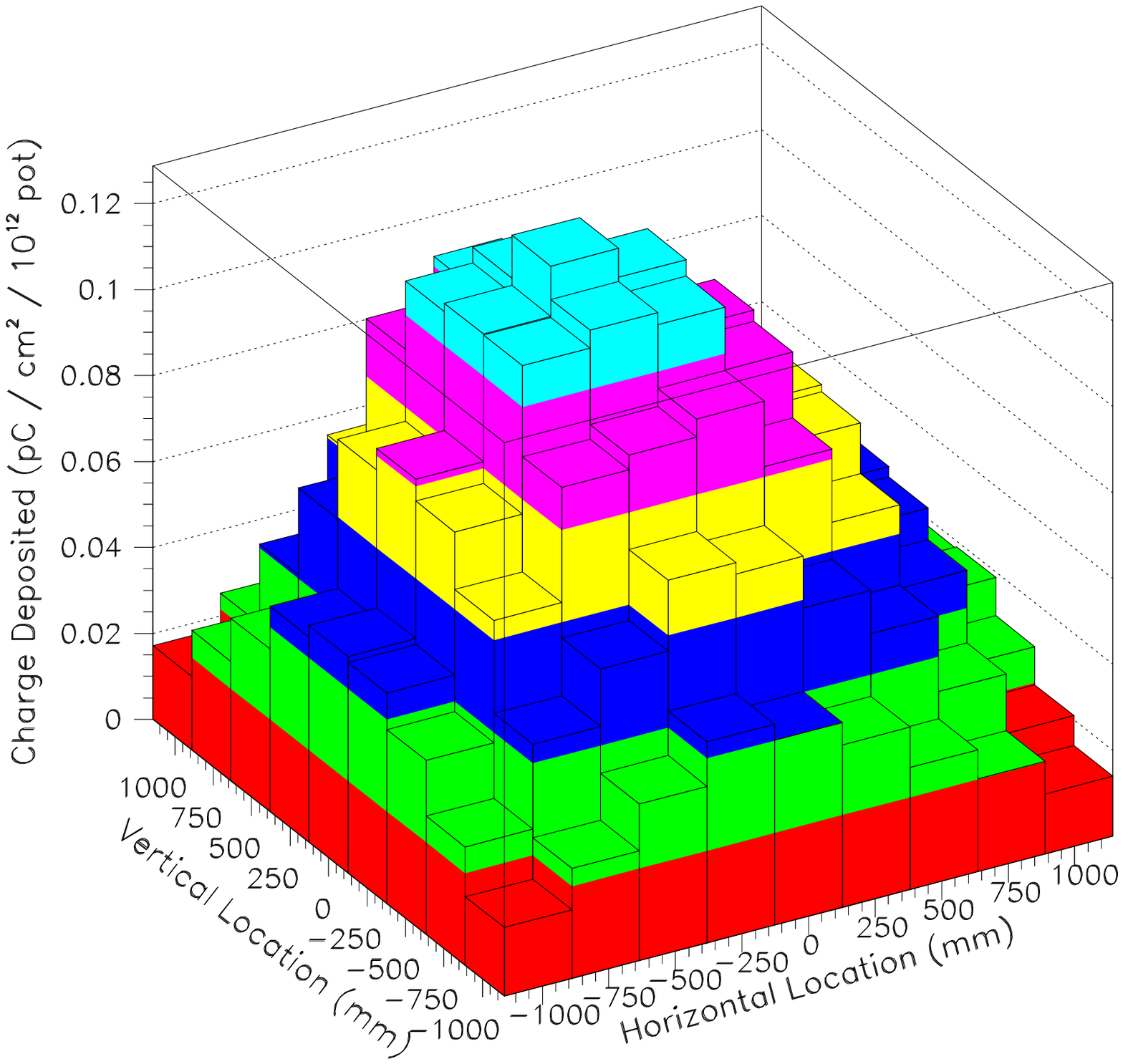}
  \hskip -.6cm
  \includegraphics[width=7cm]{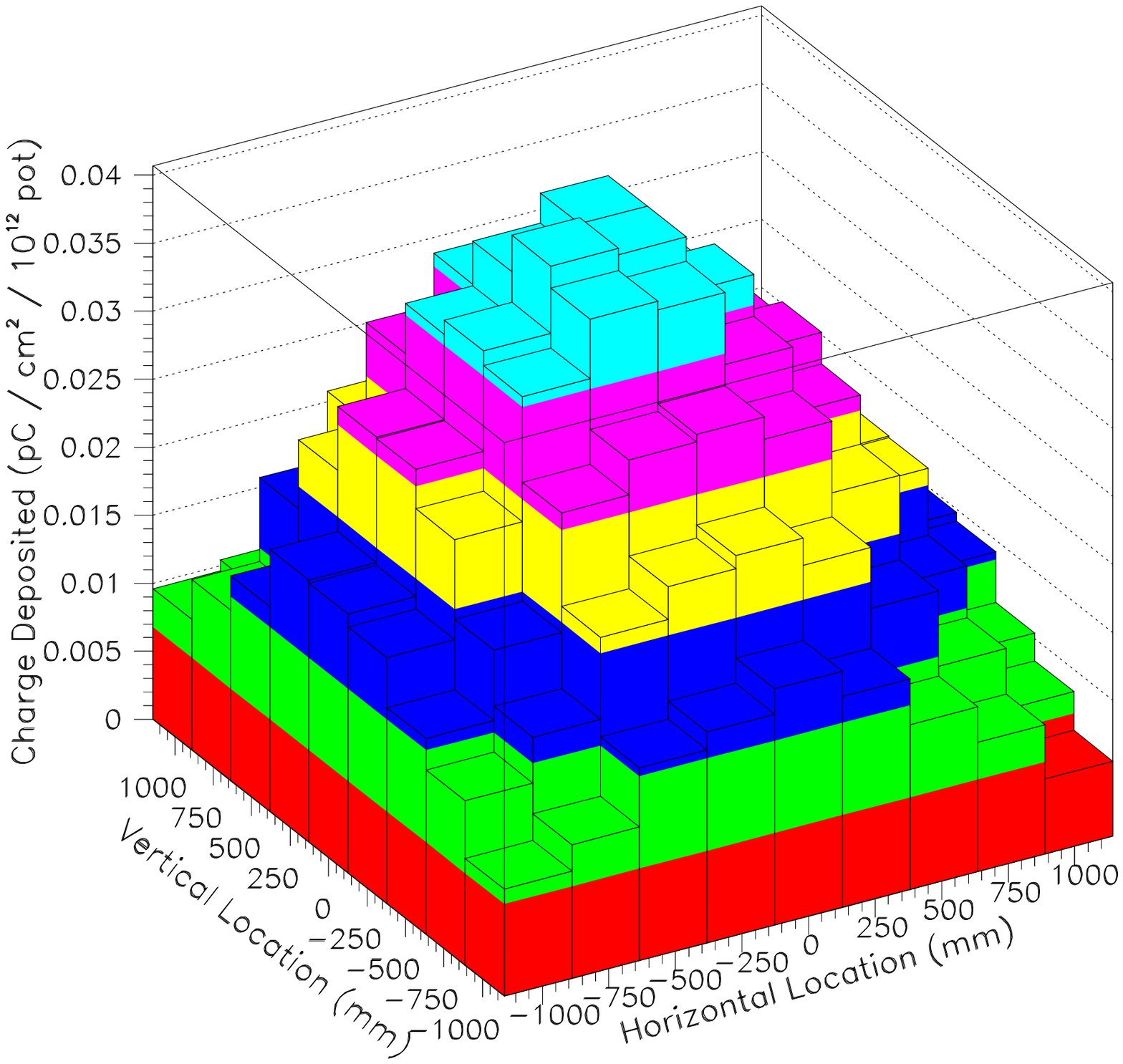}
  \vskip -.6cm
  \caption{Measured charge distributions at the second muon alcove (left column) and third muon
  alcove (right column).  The upper row shows a single beam pulse delivered to the target
  in which the horns were turned off.  The middle row shows a beam pulse in which the target
  is located in the ME position and the horns are turned on.  The lower row shows a single 
  beam pulse in which the beam, with the target in the ME position, is mis-steered into the
  upstream baffle.}
  \label{fig:perf-mu2-mu3}
\end{figure}

Figures~\ref{fig:perf-had-mu1} and \ref{fig:perf-mu2-mu3} show the data from the Hadron and Muon Monitors from three individual beam pulses.  Figure~\ref{fig:perf-had-mu1} shows the hadron monitor and muon alcove 1, while Figure~\ref{fig:perf-mu2-mu3} shows the second and third alcoves.  

The first row of these two figures shows a beam pulse in which the proton beam is centered on the NuMI target, but the horns are turned off.  The distribution of the proton beam at the Hadron Monitor is broadened by multiple Coulomb scattering in the 4~radiation length NuMI target; the observed 22~cm RMS beam size is consistent with the expected proton divergence of 0.24~mrad scattering in the target 725~m upstream.  The distribution in muon alcove 1 shows the effect of gaps in the absorber stacking, while the downstream alcoves show a lower, more peaked flux.

The second row of these two figures shows a beam pulse with the proton beam centered on target and the horns are turned on.  The target is in the Medium Energy (ME) position.
The muon fluxes are notably higher, especially for the lower-momentum muons, than the horn-off case.  The muon profiles are somewhat narrower in lateral size (and even more so for the HE position).

The third row in these two figures shows the particle distributions when the proton beam is deliberately mis-steered and impinged on the upstream collimating baffle.  This graphite baffle has a 5.5~mm radius bore.  The beam was mis-steered 2.0~mm into the baffle's volume on the right ($x>0$~mm) side of the baffle.  In this pulse, one sees the greater attenuation and scattering of the proton beam in the Hadron Monitor, and the Muon Monitor profiles show large asymmetries and increased fluences in the downstream alcoves.  The latter is due to the baffle being upstream of the target, thus acting like a target in the higher-energy target location.

\subsection{Linearity with Particle Fluence}
\label{perf_bias}

The ionization chambers were designed such that their performance would be linear at the intensities in the NuMI beam, and beam tests performed at the Fermilab Booster \cite{rdf} and the BNL ATF \cite{atf} demonstrated that charge recombination losses, potentially exacerbated by space charge build-up, is under control for the intensities envisioned in the NuMI beam.  Here, we evaluate the chambers' performance {\it in situ} by considering the central chamber of the Hadron Monitor and the central chamber of Muon Monitor alcove 1.  To date, the maximum ionization measured in a plateau curve at the Hadron Monitor is 4.2$\times10^{9}$ ionizations/cm$^3$/$\mu$s;  the maximum ionization rate observed in the Muon Monitors is 1.8$\times10^7$ ionizations/cm$^3$/$\mu$s.

\begin{figure}[!]
  \centering
  \includegraphics[width=10cm]{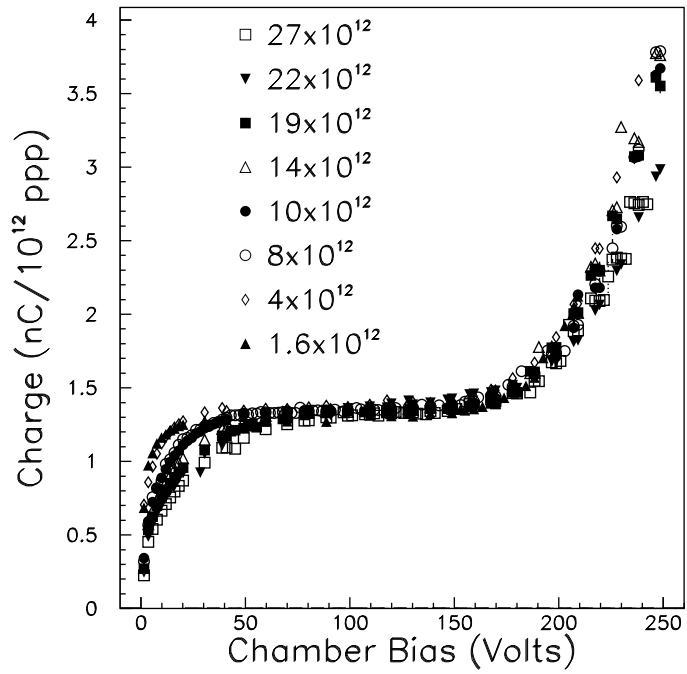}
 \vskip -.3cm
   \includegraphics[width=10cm]{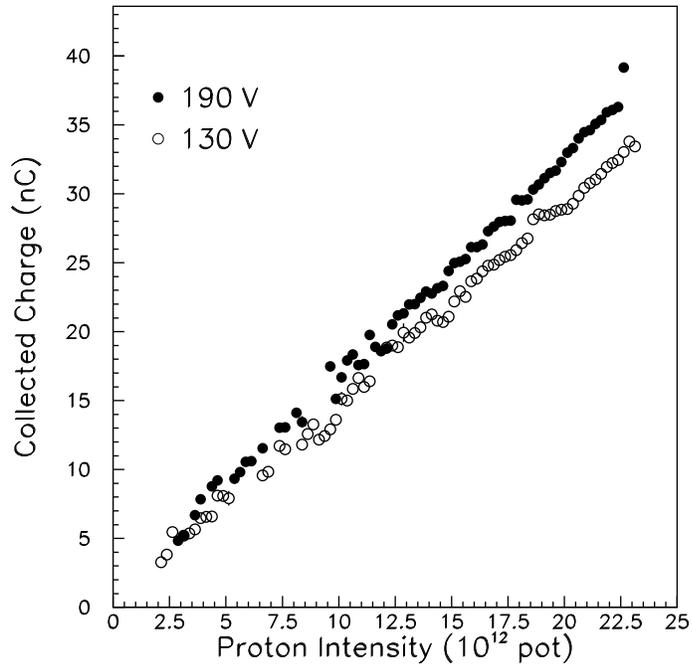}
  \vskip -.3cm
  \caption{(top)  Plateau curves of the central chamber of the Hadron Monitor for
    several different proton beam intensities. (bottom) Graph of collected charge of the
    central chamber in the Hadron Monitor as a function of the proton beam intensity.}
  \label{perf_plat_had}
\end{figure}

The top plot of Figure~\ref{perf_plat_had} shows the bias voltage curves taken with the Hadron Monitor central chamber at several beam intensities. The bias necessary to reach full collection efficiency increases with intensity (i.e. the voltage plateau is depleted). The loss of plateau is greater than would expected from the simulation performed in Ref. \cite{zwaska}, likely due to the $\approx$80~p.p.m Oxygen level in the Hadron Monitor.  Extrapolating from the plateau depletion at this intensity we expect good linearity to probably another
factor of two increase in intensity. Based on these data we have operated the
Hadron Monitor at 130 V.

\begin{figure}[!]
  \centering
  \includegraphics[width=10cm]{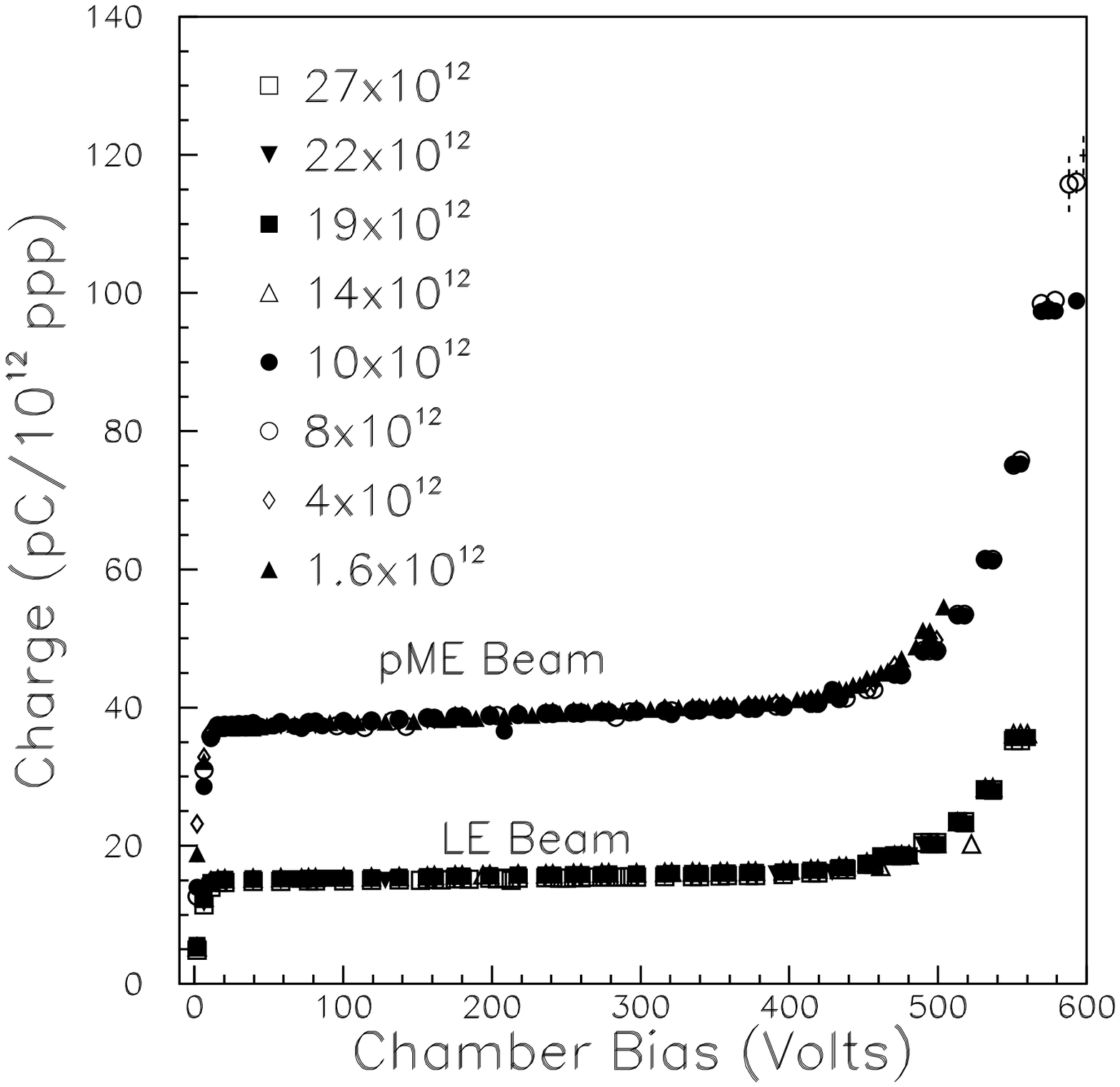}
  \vskip -.3cm
  \includegraphics[width=10cm]{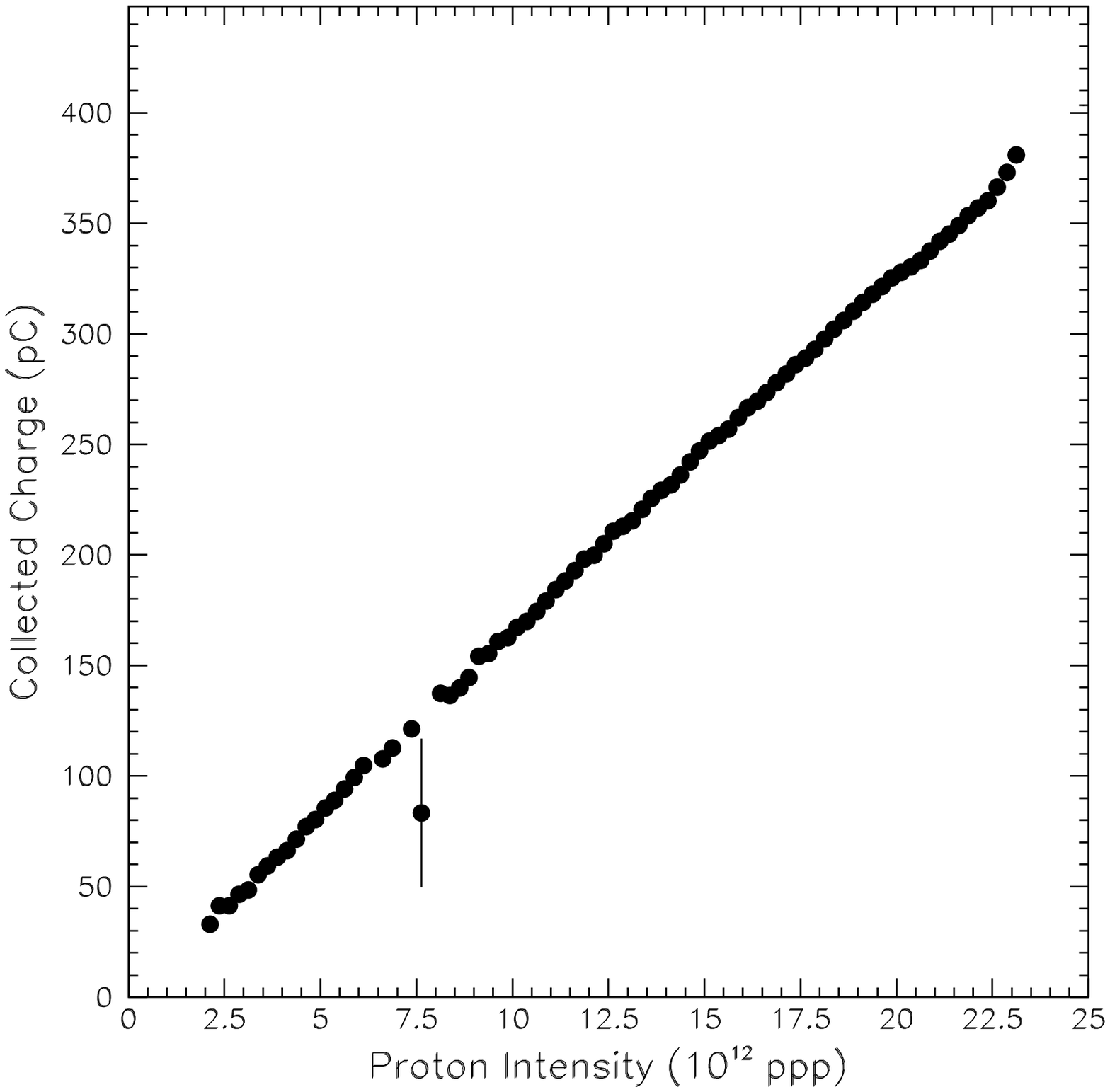}
  \vskip -.3cm
  \caption{(top)  Plateau curves of the central chamber of Muon Alcove 1 for
    several different proton beam intensities. (bottom) Graph of collected charge of the
    central chamber in Muon Alcove 1 as a function of the proton beam intensity.}
  \label{perf_plat_mu}
\end{figure}

The top plot of Figure~\ref{perf_plat_mu} shows the bias voltage curves for the Muon Monitor
alcove 1 central pixel. It reaches complete collection
efficiency at $\approx$15 V.   Based on this meager depletion, we do not
expect any problems until 100$\times$ the highest muon intensity 
reached in the NuMI beam.  Based on these data, we have operated 
the Muon Monitors at 300~V.

To study the chambers' linearity {\it vs.} particle fluence, we studied the data accumulated over several months that fills in much of the desired range of intensities.  The collected charge for the central chamber in the Hadron Monitor is shown as a function of proton beam intensity in the lower plot of Figure~\ref{perf_plat_had}.  The response is linear, with the 130V data having a response of $\approx$1.46~nC/$10^{12}$~ppp.  The 190~V data has a slope of $\approx$1.64~nC/$10^{12}$~ppp.  The ratio of these two slopes is 1.12, consistent with expectations from the plateau curves of Figure~\ref{perf_plat_had}.  The scatter in these curves is coincident with periods of varying proton beam conditions (spot size, location on target) as recorded by the pre-target foil SEM.  

The collected charge for the central pixel of Muon Alcove 1 is shown in the lower plot of Figure~\ref{perf_plat_mu}.  As above with the Hadron Monitor, the response is mostly linear.  The muon signal sustains the same temporal variability as the Hadron Monitor signal, further suggesting that the variation is because of beam conditions and not chamber response.

\subsection{Chamber Stability}

The ionization chambers must provide a stable response to particle fluences over long periods of time in order to track changes in beam quality.  Because the chambers are under gas flow with the gas exhausted to the room, changes in barometric pressure, changes in flow that result in pressure changes within the chambers, or changes in temperature can affect the ionization chamber response, which is expected to grow linearly with gas density.  As mentioned in Section~\ref{beamon_gas}, we installed absolute pressure monitors and temperature RTDs on each chamber array to track these variables so that the chamber responses can be corrected.  As shown in Section~\ref{calib}, such corrections were shown on the bench to maintain stability at the 1\% level over a period of one year.

\begin{figure}[!]
  \centering
  \includegraphics[width=15cm]{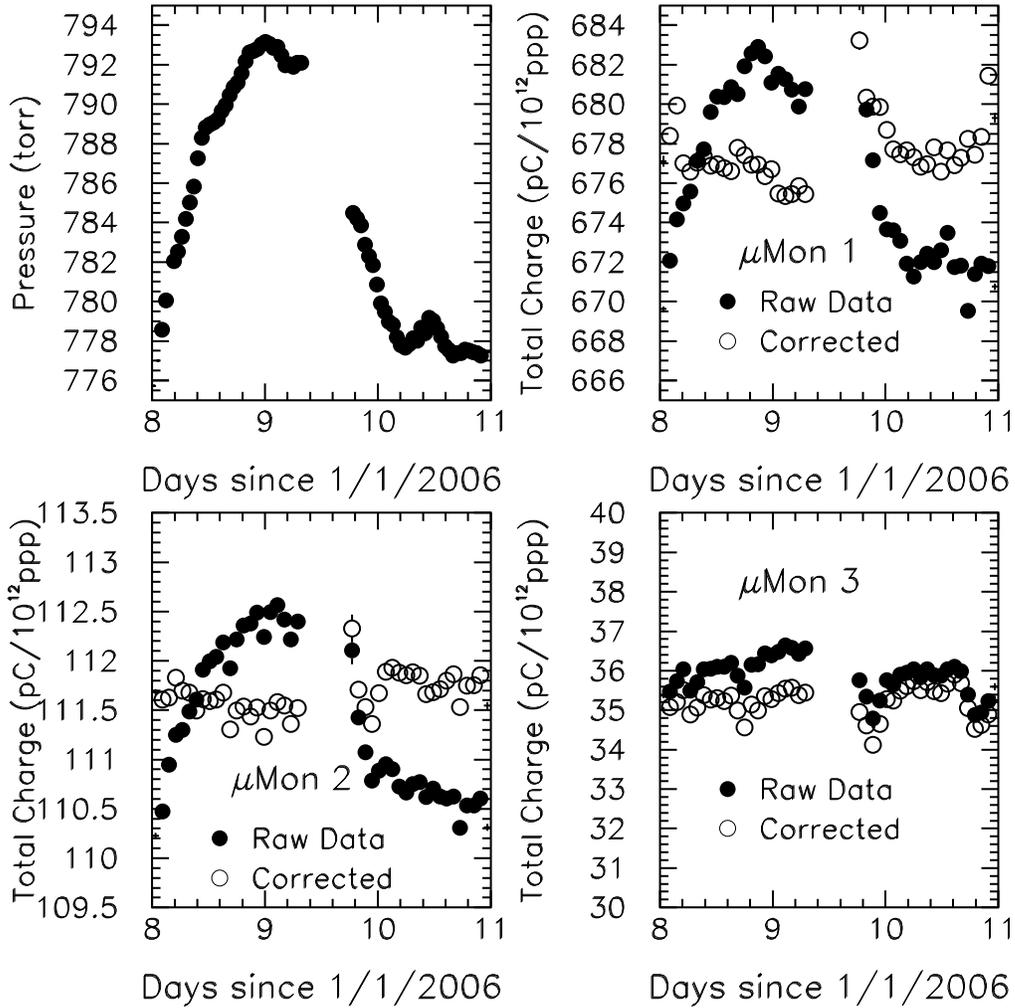}
  \vskip -.3cm
  \caption{Top left plot shows the pressure in the Hadron Monitor over a three day period (Muon array pressures are similar).  Remaining plots show the response of Muon Alcoves $1-3$ during this time, before and after the temperature and pressure corrections of Section~\ref{calib}. }
  \label{fig:mmon_stability1}
\end{figure}

\begin{figure}[!]
  \centering
  \includegraphics[width=15cm]{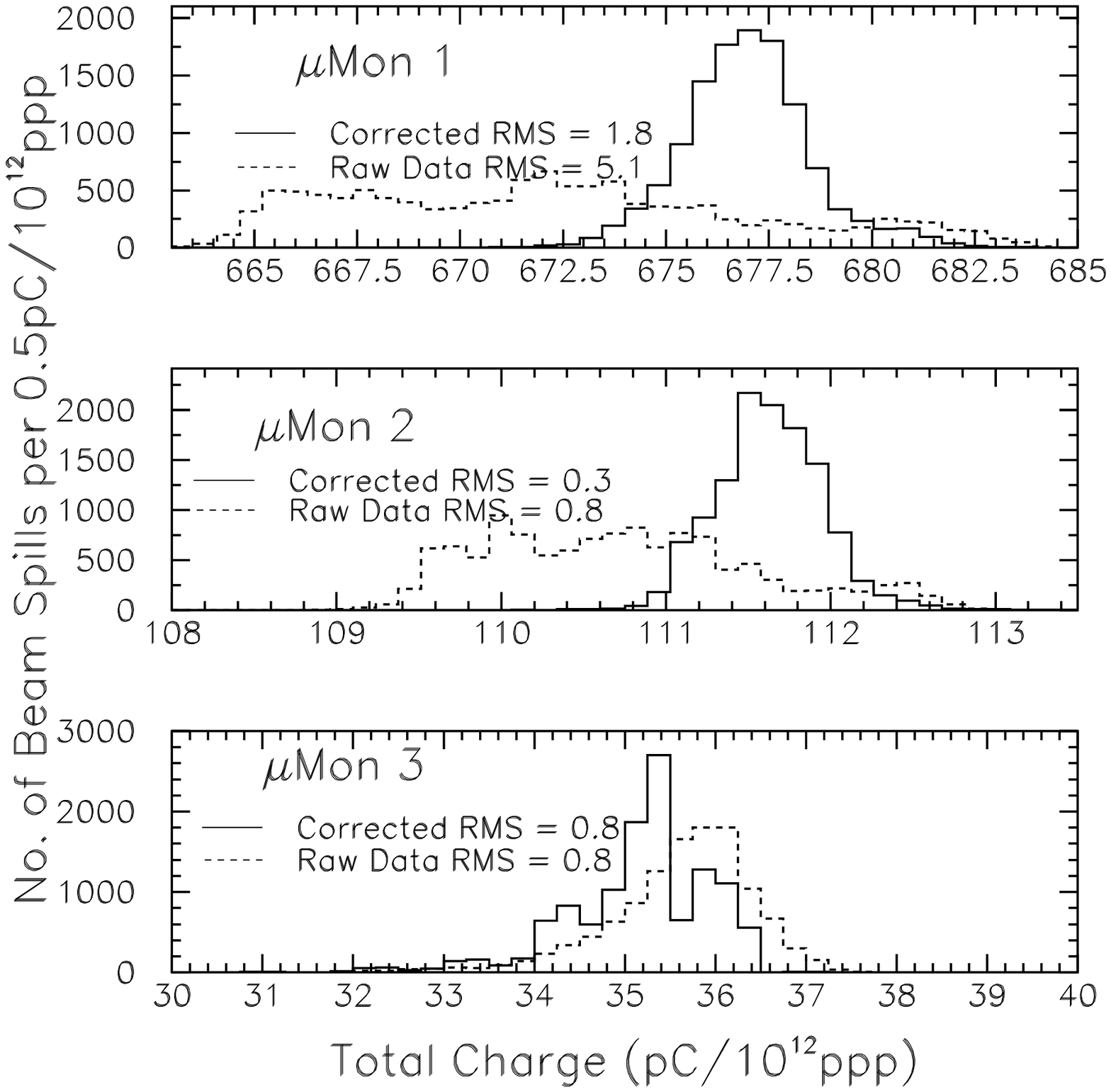}
  \vskip -.3cm
  \caption{Projections of the chamber responses plotted in Figure~\ref{fig:mmon_stability1}, before and after the temperature and pressure corrections of Section~\ref{calib}. }
  \label{fig:mmon_stability2}
\end{figure}

The chamber response can be calibrated {\it in situ} in the beam.  Shown in Figure~\ref{fig:mmon_stability1} is the absolute pressure in the Hadron Monitor array over a three day period in the run in which the pressure changes by $\sim20$~Torr.  As can be seen in Figure~\ref{fig:mmon_stability2}, the chamber responses, normalized to recorded charge per proton on target, is constant to 1\% over the several day period.\footnote{Figure~\ref{fig:mmon_stability1} shows a $>1\%$ effect when beam resumed after a half-day shutdown:  temperatures in the target station cooled off, decreasing the resistance in the horns and increasing the current delivered to the horns.  The changes in these quantities, monitored independently, are corroborated by the excursions in muon fluxes after beam startup.}  The corrections normalize the chamber response to that expected at 20$^\circ$C and 790~Torr, so consequently the mean charge is shifted after the corrections for this particular period of the run in which the typical pressure was closer to 780~Torr.

\section{Beam Measurements}
\label{beam-meas}

The secondary and tertiary beam 
monitors have important roles in: (a) confirmation of the neutrino
flux predictions from the beam Monte Carlo, and (b) monitoring
the quality of the secondary and tertiary beams.  A number of studies
were undertaken to demonstrate the monitors' sensitivity to the
horn optics, proton beam steering on the target, etc. The studies confirm the measurement capabilities of 
the monitors.  Additionally, we describe how the monitors were
used as a diagnostic tool during the failure of the NuMI
target.

\subsection{Sensitivity to Proton Beam Position}
\label{bmeas_muons_targ}

The Muon Monitors are quite sensitive to changes in the steering
of the proton beam onto the target.  Thus, this system can
be used as an independent check of the proton beam steering, 
complementing the extrapolation provided by the primary beam 
instrumentation.   We performed several scans early in the NuMI 
commissioning process which can be used to calibrate any 
subsequent excursions detected by the Muon Monitors.  The results of
this study have been confirmed with two accident spills during subsequent running, one in March, 2005, and one
in November, 2005.  

\begin{figure}[!]
  \centering
  \vskip -1.cm
  \includegraphics[width=15cm]{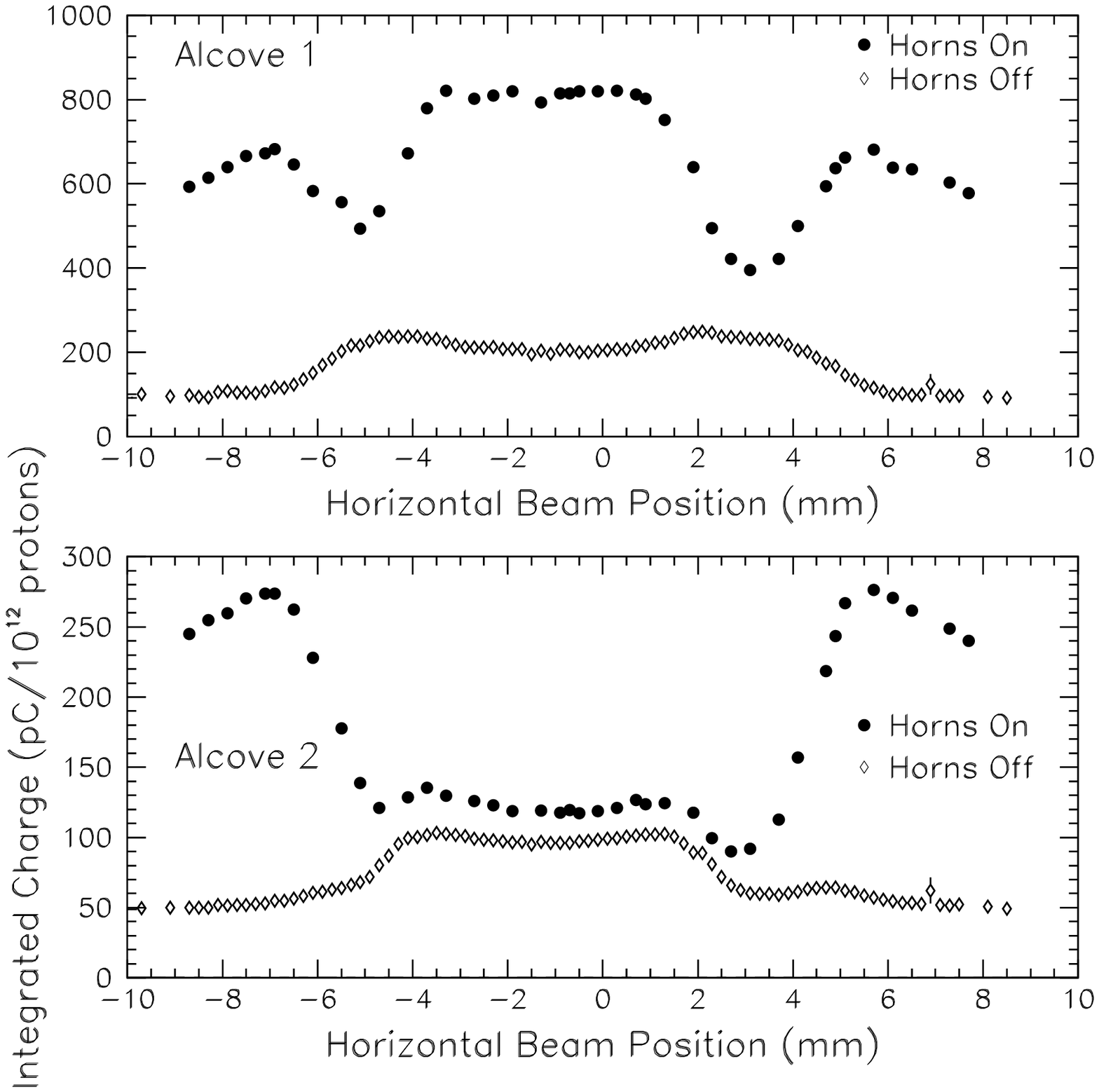}
  \vskip -.7cm
  \includegraphics[width=15cm]{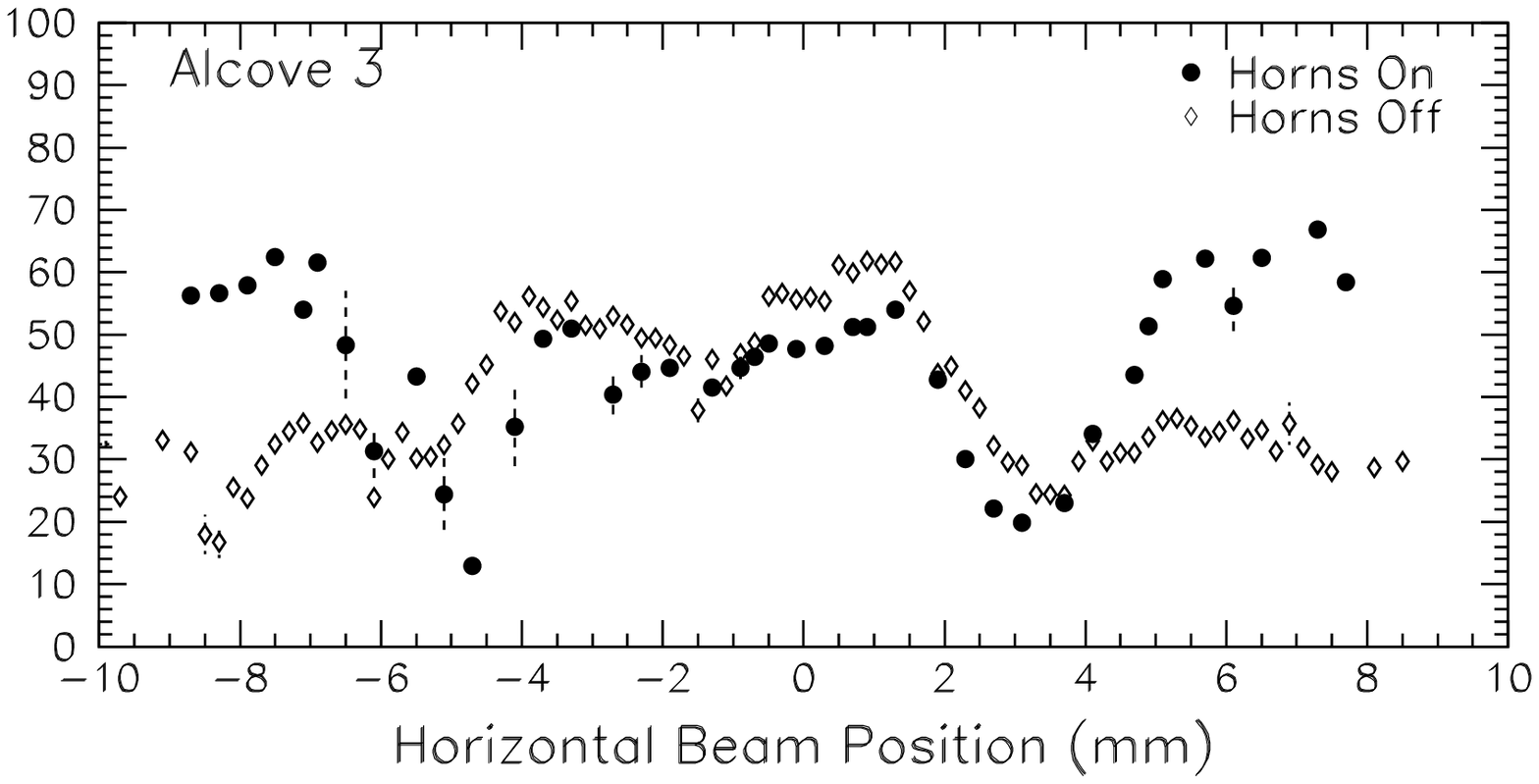}
  \vskip -.5cm
  \caption{Muon Monitor results from horizontal target scans in the
    low-energy position.  The target is centered at -0.95 mm, and its edges are at $+2.0$~mm and $-4.4$~mm.  The inner aperture of the baffle is at $-6.5$~mm and $+4.5$~mm.}
  \label{bmeas_hmuons_le}
\end{figure}

\begin{figure}[!]
  \centering
  \vskip -1.cm
  \includegraphics[width=15cm]{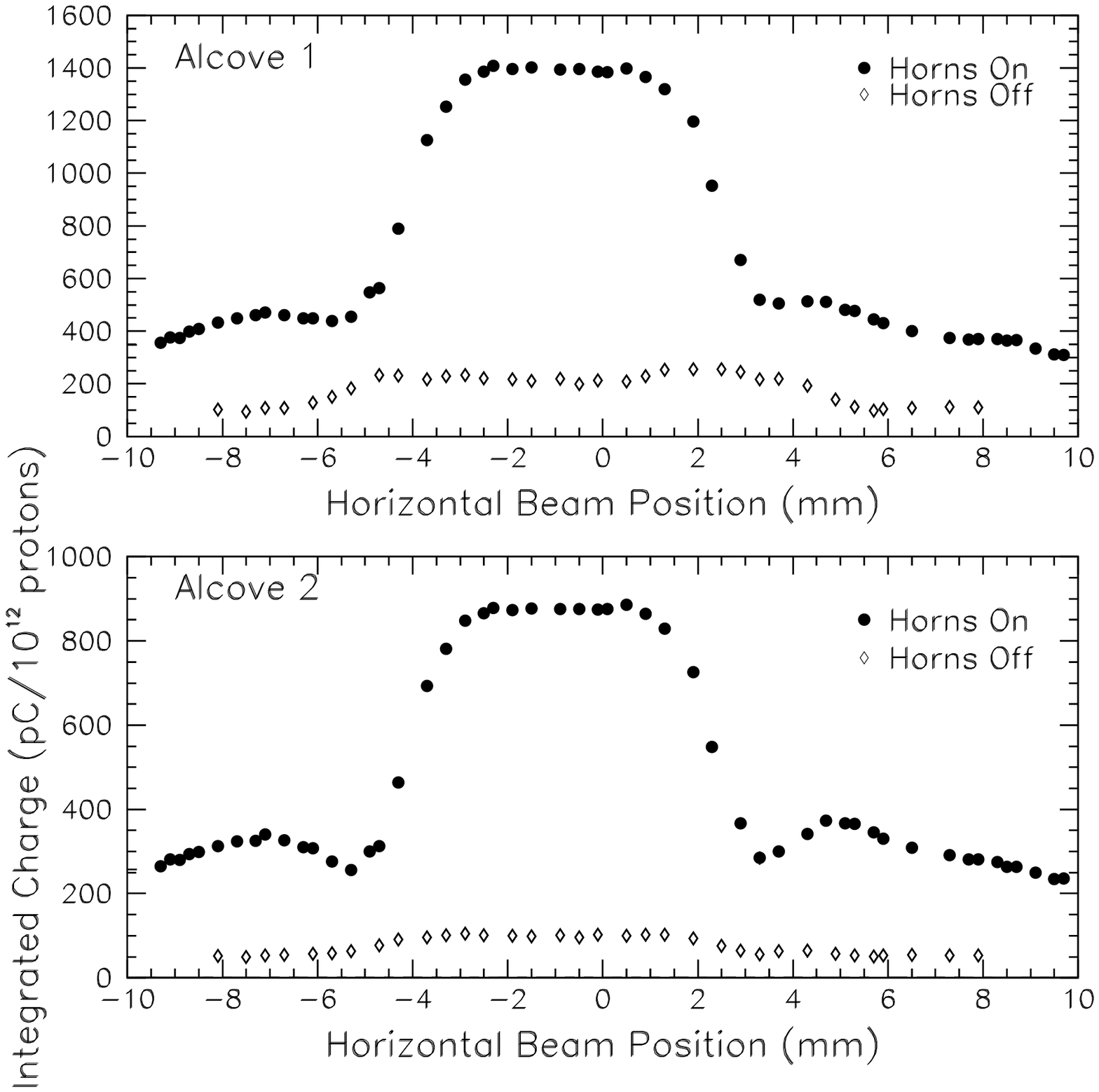}
  \vskip -.7cm
  \includegraphics[width=15cm]{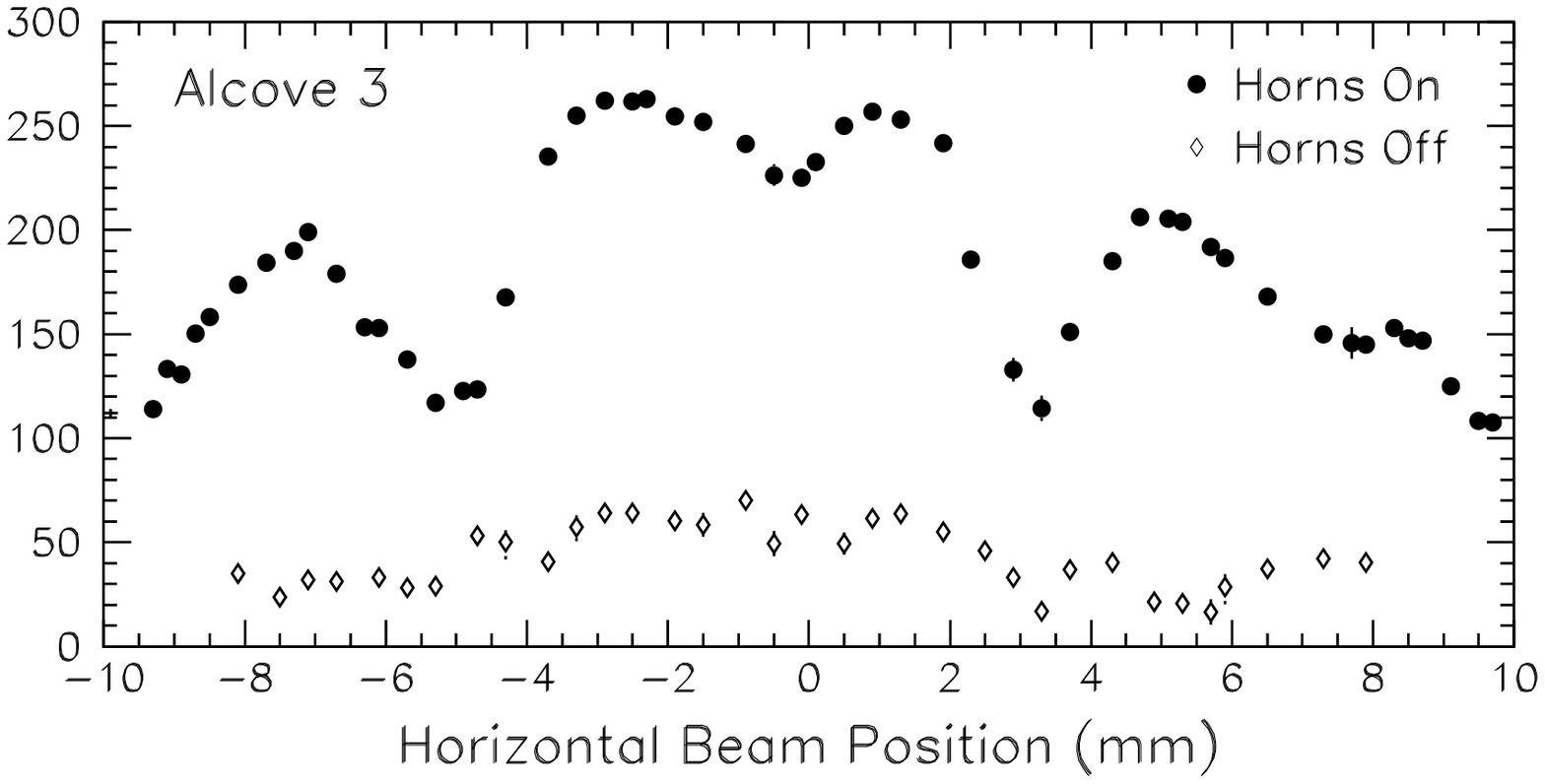}
  \vskip -.5cm
  \caption{Muon Monitor results from horizontal target scans with the target in the high-energy position.  The target is centered at -0.95 mm, and its edges are at 
     $+2.0$~mm and $-4.4$~mm.  The inner aperture of the baffle is at $-6.5$~mm and $+4.5$~mm.}
  \label{bmeas_hmuons_he}
\end{figure}

The summed signals from the muon alcoves are shown in Figure~\ref{bmeas_hmuons_le} and Figure~\ref{bmeas_hmuons_he} for two scans of the proton beam across the target with the beam in the low-energy configuration and high-energy configurations.\footnote{The alcove 1 signal comes from both the decay of pions and from particles emanating from the Hadron Absorber.  Its peaks in intensity, without horn focusing, correspond to when the proton beam penetrates in the region between the baffle and target and strikes the absorber.}  The signals in all 3 alcoves display prominently the edges of the target at $+2.0$~mm and $-4.4$~mm, and the inner aperture of the upstream collimating baffle at $-6.5$~mm and $+4.5$~mm.  The flux of higher energy muons (alcoves 2 \& 3) rises noticeably when the proton beam strikes near the edge of the target:  high energy pions have a lesser probability of reinteracting in the target when created at the edge.    
Other experiments have used their Muon Monitors more extensively to align the target hall components (see \cite{Astier,Casagrande}).  In that case the proton beam and components were moved until the muon flux was maximized; such a method cannot be used for the NuMI alignment.  As can be seen in Figures~\ref{bmeas_hmuons_le}, \ref{bmeas_hmuons_he}, centering the proton beam on the target does not maximize muon flux, owing to pion reinteractions in the target which deplete the number of medium- and high-energy pions.  We therefore do not align our beam components by maximizing the muon yield, and instead developed an independent means of beam-based alignment of the beam components, 
which will be described in a forthcoming article \cite{bba}.  As a result of this alignment, the proton beam is steered at $-1.2$~mm.

\begin{sidewaysfigure}
  \centering
  \includegraphics[width=7.5cm]{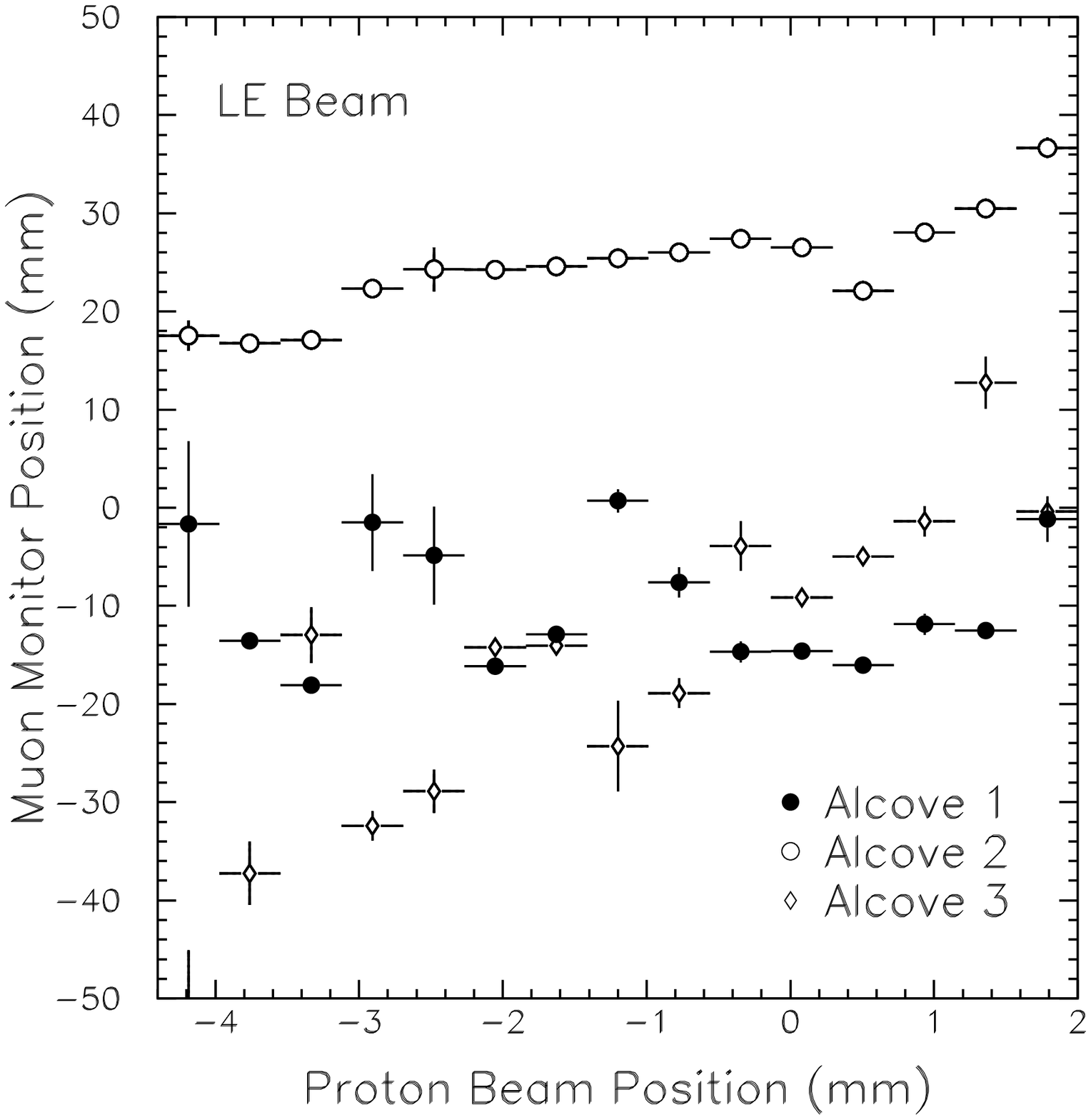}
  \includegraphics[width=7.5cm]{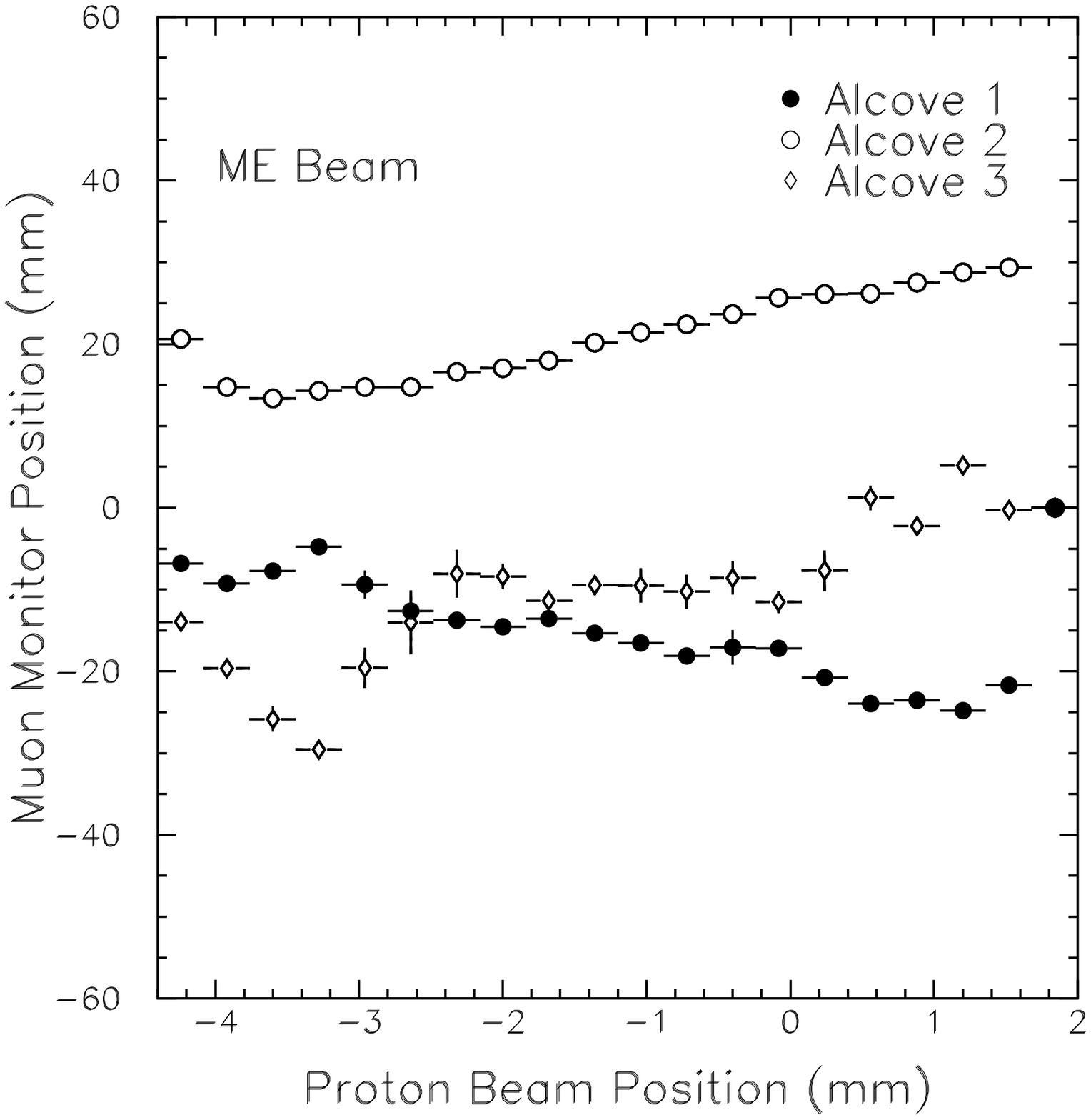}
  \includegraphics[width=7.5cm]{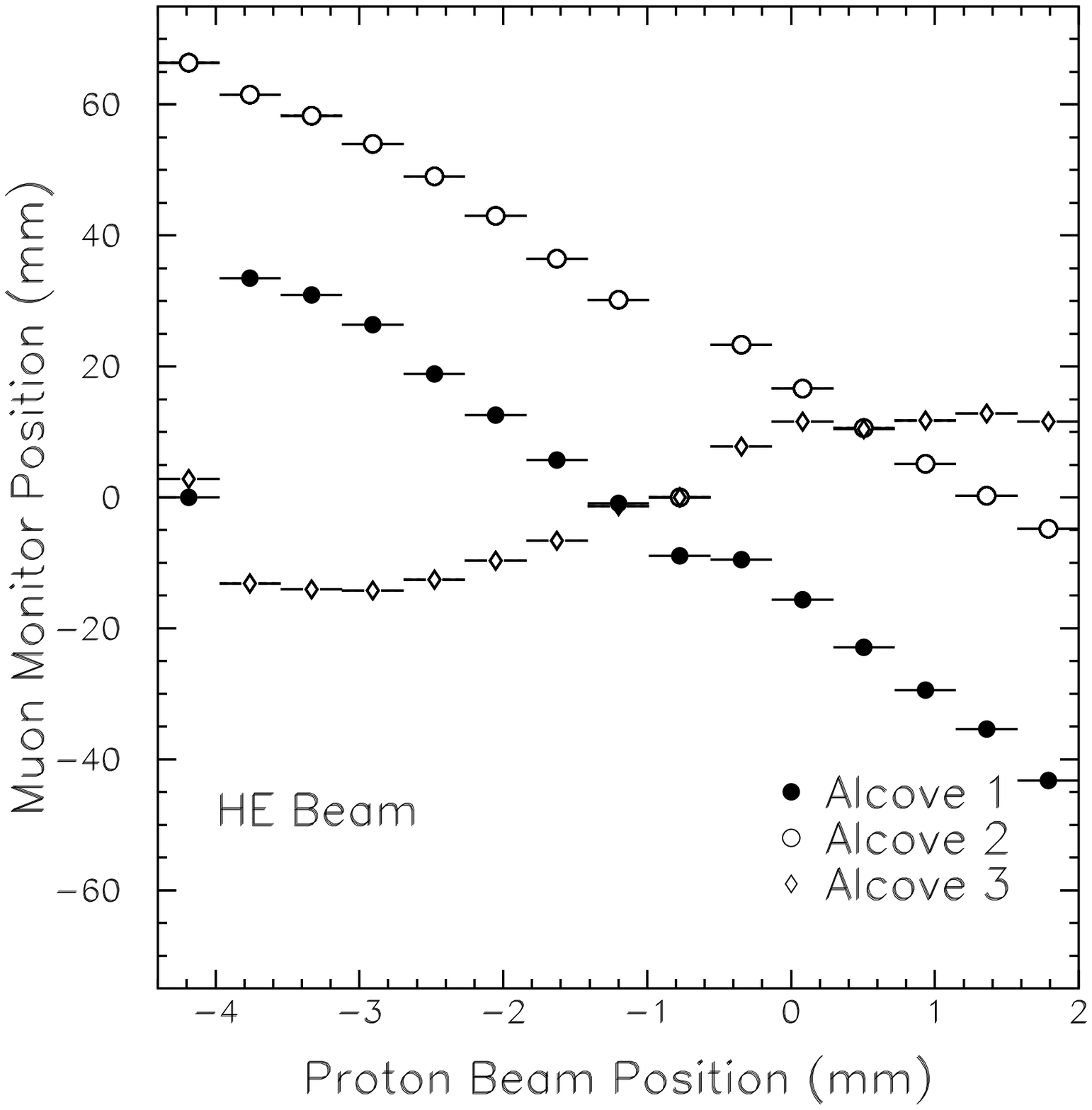}
  \caption{Muon Monitor centroids during scans of the proton beam across the target in the various
    beam configurations.  }
  \label{bmeas_hmuons_hpos}
\end{sidewaysfigure}

The centroid of the distributions measured at the Muon Monitors is capable of confirming the proper pointing of the proton beam.  Figure~\ref{bmeas_hmuons_hpos} shows the horizontal centroid position of the muon profiles measured during horizontal scans of the proton beam across the target.  The horn focusing acts like a lens for various energies of pions and can overfocus or underfocus them\footnote{In some cases the pions are not focused at all, because they pass through the neck of the horn.  However, they still acquire an angle if steered away from the center of the target.}.  The centroid
positions measured in the alcoves thus are correlated (positively or negatively) with the beam position on target, the proportion being dependent on the focusing.  As each alcove samples a different portion of the muon energy spectrum, the alcoves' centroid correlation will vary differently.  From these distributions, it is clear that the Muon Monitors can
detect future excursions of $\approx1$~mm of the proton beam off target center.\footnote{It should be noted that the primary beam instrumentation has been designed to measure the proton beam at the target within 50~$\mu$m.}

\subsection{Target Integrity}
\label{bmeas_water}

\begin{figure}[t]
  \centering
  \includegraphics[width=13.cm]{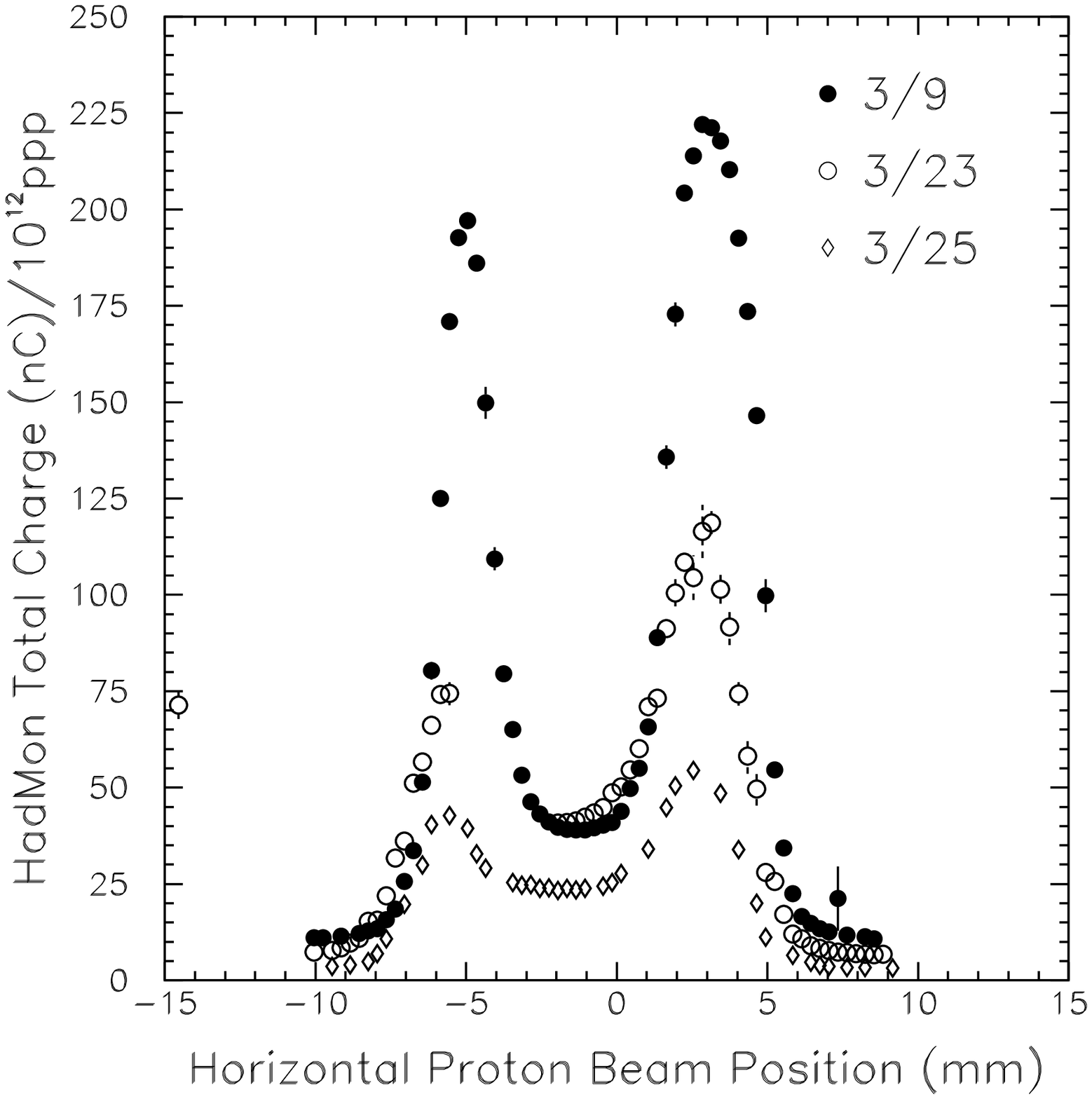}
  \vskip -.1cm
  \caption{Hadron Monitor normalized intensity of horizontal 
    target scans at several dates during the
    target incident.  The March 9 data show a typical horizontal scan,
    presumably before any damage.  The March 23 and 28 data show reduced
    signal in the Hadron Monitor from water in the target canister. }
  \label{fig:bmeas_water_dia_hint}
\end{figure}

On March 23, 2005 water was found in the vacuum pump keeping the NuMI target canister evacuated.  Such an effect could arise if a leak develops in the water lines cooling the target's graphite fins.  Water leaking from the cooling lines can fill the vacuum vessel, rendering the target inoperable because of the concern for thermal expansion of the water in the vessel.
Several scans of the proton beam\footnote{The proton beam spot at this time was $\sigma_x\times\sigma_y=0.7\times1.4$~mm$^2$.} across the target were performed, the data from which are shown in Figure~\ref{fig:bmeas_water_dia_hint}.  Under normal circumstances, the total charge seen at the Hadron Monitor should drop if the proton beam is directed at the target or at the upstream collimating baffle, both of which are 2-3 interaction lengths of graphite.  The beam charge arriving at the Hadron Monitor should increase when the proton beam is directed at a gap between the inner aperture of the baffle and the outer edges of the target.  

\begin{figure}[t]
  \centering
  \includegraphics[width=13cm]{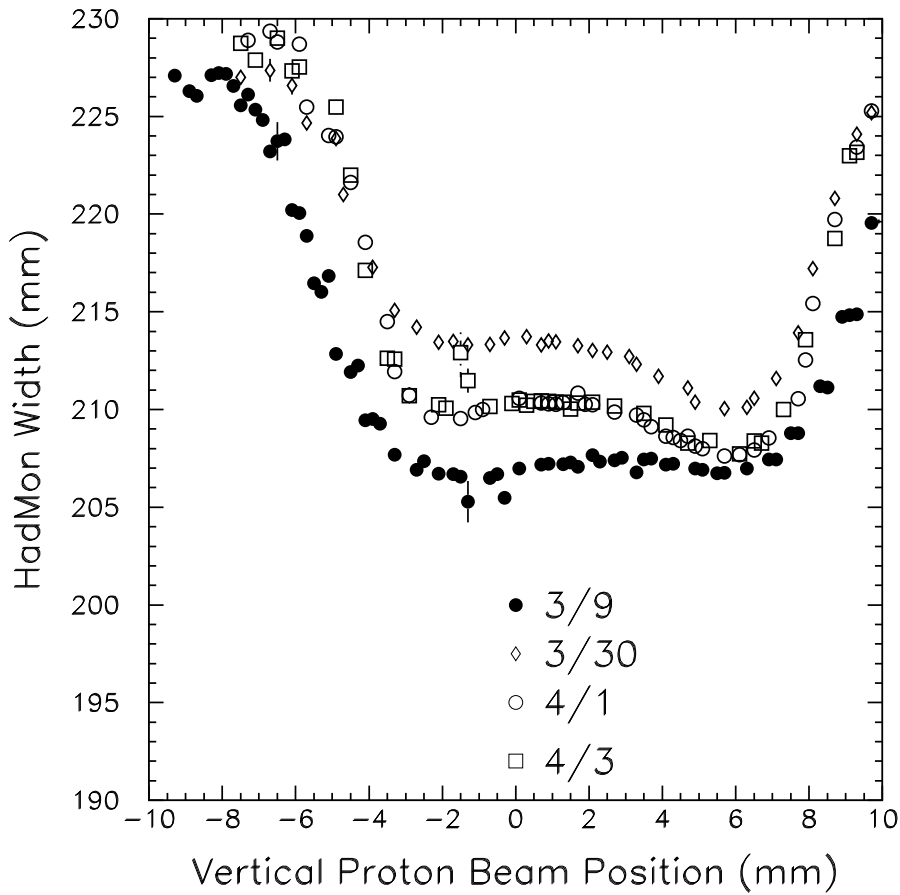}
  \vskip -.1cm
  \caption{Measurement of the RMS beams size at the Hadron Monitor location as a function of the vertical location of the proton beam at the target.  The beam was translated vertically, so as to scan it through the edges of the collimating baffle and target center fin.  The increased beam size after March 23 indicates the presence and level of water in the target canister.}
  \label{bmeas_water_dia_vsig}
\end{figure}

The beam scans on March 23 and following showed an attenuation of the proton beam passing through the target-baffle gap, indicating the presence of additional inert material.  In fact, the 1.6~interaction lengths inferred from the March~25 scan corresponds to 1.3~m of water, indicating that water filled the entire length of the target vacuum vessel.  

It was possible to assess the level of the water within the vacuum vessel by means of a vertical scan of the proton beam across the target.  Such a set of scans is shown in Figure~\ref{bmeas_water_dia_vsig}, which shows the RMS size of the beam arriving at the Hadron Monitor as a function of the proton beam's vertical position at the target.  For the scan dated March~9, prior to the accident, one notes a sharp rise in the beam size at $-6$~mm and $+6$~mm due to the proton beam scattering in the upstream collimating baffle, as well as a small amount of additional scattering at the target center due to one 0.2$X_0$ graphite fin which is narrow in this view.  

After the accident, however, one notes an increase in the Hadron Monitor beam size as the beam is scanned downward into the target vessel.  The level of this increase was used to determine the water level in the canister, and to pinpoint the likely location of the leak at one of the water cooling lines in the vessel.  Attempts on April 1 and 3 to purge the water out of the vessel by overpressuring the vessel with Helium gas were only partially successful, as determined by the data from the Hadron Monitor.

Ultimately, the target assembly was removed from the beamline.  Inspection revealed that the water inside the target canister
was consistent with the water level indicated with the Hadron Monitor data.  The target was drained and overpressured by 1.6~atm of He gas to forestall any further leaks of the cooling system.  As a means of regularly assessing the target's integrity during subsequent beam operations, the proton beam is scanned across its surface at one-month intervals, and the data from the Hadron Monitor utilized to detect any water leakage.

\section{Conclusions}

We have built ionization chambers to monitor the secondary and tertiary beams of the NuMI neutrino facility at FNAL.  The intense charged particle fluences in the beam line have not proven to limit the chambers operation;  space charge build-up within the ion chamber results in negligible charge loss due to recombination, consistent with expectations from earlier beam tests.  The chambers have proven robust in the intense radiation field near the beam stop.  The diagnostic capabilities of the secondary beam system provide redundant checks of the proton beam's stability on target, as well as of the integrity of the target and focusing horns which produce the secondary beam.  


\section{Acknowledgements}
It is a pleasure to thank S. O'Kelley of the University of Texas Nuclear Engineering Teaching Center, the staff of the University of Texas Physics Department Mechanical Shops, K. Lang and T. Tipping of the University of Texas, K. Kriesel of the University of Wisconsin Physical Sciences Laboratory, and B. Baller, D. Bogert, R. Ducar, J. Hylen, D. Pushka, W. Smart, G. Tassotto, and K. Vaziri of Fermilab for valuable collaboration on this project.  This work was supported by the U.S. Department of Energy under contracts DE-FG03-93ER40757,  DE-FG02-95ER40896 and DE-AC02-76CH3000, and by the University of Texas at Austin Fondren Family Foundation.

\end{document}